\documentclass{article}
\usepackage{amsfonts}
\usepackage{amssymb}
\usepackage[dvips]{color}
\usepackage{graphicx}
\usepackage{amsmath}

\newtheorem{theorem}{Theorem}

\newtheorem{corollary}[theorem]{Corollary}

\newtheorem{definition}[theorem]{Definition}

\newtheorem{lemma}[theorem]{Lemma}

\newtheorem{proposition}[theorem]{Proposition}
\newtheorem{remark}[theorem]{Remark}

\newenvironment{proof}[1][Proof]{\textbf{#1.} }{\ \rule{0.5em}{0.5em}}

\input{tcilatex}
\begin{document}

\title{Dual Representation of Quasiconvex Conditional Maps}
\author{Marco Frittelli\thanks{%
Dipartimento di Matematica, Universit\`{a} degli Studi di Milano.} \and %
Marco Maggis\thanks{%
Dipartimento di Matematica Universit\`{a} degli Studi di Milano.}}
\date{January 20, 2010}
\maketitle

\begin{abstract}
We provide a dual representation of quasiconvex maps $\pi :L_{\mathcal{F}%
}\rightarrow L_{\mathcal{G}}$, between two lattices of random variables, in
terms of conditional expectations. This generalizes the dual representation
of quasiconvex real valued functions $\pi :L_{\mathcal{F}}\rightarrow 
\mathbb{R}$ and the dual representation of conditional convex maps $\pi :L_{%
\mathcal{F}}\rightarrow L_{\mathcal{G}}.$
\end{abstract}



\section{Introduction\protect\footnote{\textbf{Acknowledgements} We wish to
thank Dott. G. Aletti for helpful discussion on this subject.}}

Quasiconvex analysis has important applications in several optimization
problems in science, economics and in finance, where convexity may be lost
due to absence of global risk aversion, as for example in Prospect Theory 
\cite{KT}.

The first relevant mathematical findings on quasiconvex functions were
provided by De Finetti \cite{DeFin} and since then many authors, as \cite{Fe}%
, \cite{Cr77}, \cite{Cr80}, \cite{ML81}, \cite{PP84} and \cite{PV} - to
mention just a few, contributed significantly to the subject. More recently,
a Decision Theory complete duality involving quasiconvex real valued
functions has been proposed by \cite{CMMMb}. For a review of quasiconvex
analysis and its application and for an exhaustive list of references on
this topic we refer to Penot \cite{P}.

A function $f:L\rightarrow \overline{\mathbb{R}}:=\mathbb{R}\cup \left\{
-\infty \right\} \cup \left\{ \infty \right\} $ defined on a vector space $L$
is quasiconvex if for all $c\in \mathbb{R}$ the lower level sets $\left\{
X\in L\mid f(X)\leq c\right\} $ are convex. In a general setting, the dual
representation of such functions was shown by Penot and Volle \cite{PV}. The
following theorem, reformulated in order to be compared to our results, was
proved by Volle \cite{Volle}, Th. 3.4. As shown in the Appendix \ref{sec52},
its proof relies on a straightforward application of Hahn Banach Theorem.

\begin{theorem}[\protect\cite{Volle}]
\label{Volle1}Let $L$ be a locally convex topological vector space, $%
L^{\prime }$ be its dual space and $f:L\rightarrow \overline{\mathbb{R}}:=%
\mathbb{R}\cup \left\{ -\infty \right\} \cup \left\{ \infty \right\} $ be
quasiconvex and lower semicontinuous. Then 
\begin{equation}
f(X)=\sup_{X^{\prime }\in L^{\prime }}R(X^{\prime }(X),X^{\prime })
\label{111}
\end{equation}%
where $R:\mathbb{R\times }L^{\prime }\rightarrow \overline{\mathbb{R}}$ is
defined by 
\begin{equation*}
R(t,X^{\prime }):=\inf_{\xi \in L}\left\{ f(\xi )\mid X^{\prime }(\xi )\geq
t\right\} .
\end{equation*}
\end{theorem}

The generality of this theorem rests on the very weak assumptions made on
the domain of the function $f,$ i.e. on the space $L$. On the other hand,
the fact that only \emph{real valued} maps are admitted considerably limits
its potential applications, specially in a dynamic framework.

\bigskip

To the best of our knowledge, a \emph{conditional} version of this
representation is lacking in the literature. When $(\Omega ,\mathcal{F},(%
\mathcal{F}_{t})_{t\geq 0},\mathbb{P})$ is a filtered probability space,
many problems having dynamic features leads to the analysis of maps $\pi
:L_{t}\rightarrow L_{s}$ between the subspaces $L_{t}\subseteq L^{1}(\Omega ,%
\mathcal{F}_{t},\mathbb{P})$ and $L_{s}\subseteq L^{0}(\Omega ,\mathcal{F}%
_{s},\mathbb{P})$, $0\leq s<t$.

In this paper we consider quasiconvex maps of this form and analyze their
dual representation. We provide (see Theorem \ref{main} for the exact
statement) a conditional version of (\ref{111}):%
\begin{equation}
\pi (X)=ess\sup_{Q\in L_{t}^{\ast }\cap \mathcal{P}}R(E_{Q}[X|\mathcal{F}%
_{s}],Q),  \label{1234}
\end{equation}%
where 
\begin{equation*}
R(Y,Q):=ess\inf_{\xi \in L_{t}}\left\{ \pi (\xi )\mid E_{Q}[\xi |\mathcal{F}%
_{s}]\geq_Q Y\right\} ,\text{ }Y\in L_{s},
\end{equation*}%
$L_{t}^{\ast }$ is the order continuous dual space of $L_{t}$ and $\mathcal{P%
}=:\left\{ \frac{dQ}{d\mathbb{P}}\mid Q<<\mathbb{P}\right\} $.

\bigskip

Furthermore, we show that if the map $\pi $ is quasiconvex, monotone and
cash additive then it is very easy to derive from (\ref{1234}) the well
known representation of a conditional convex risk measure \cite{Sca}.

\bigskip

The formula (\ref{1234}) is obtained under quite weak assumptions on the
space $L_{t}$ which allow us to consider maps $\pi $ defined on the typical
spaces used in the literature in this framework: $L^{\infty }(\Omega ,%
\mathcal{F}_{t},\mathbb{P}),$ $L^{p}(\Omega ,\mathcal{F}_{t},\mathbb{P})$,
the Orlicz spaces $L^{\Psi }(\Omega ,\mathcal{F}_{t},\mathbb{P}).$

We state our results under the assumption that $\pi $ is lower
semicontinuous with respect to the weak topology $\sigma (L_{t},L_{t}^{\ast
}).$ As shown in Proposition \ref{LemmaCFB} this condition is equivalent to
continuity from below, which is the natural requirement in this context.

\bigskip

The proof of our main Theorem \ref{main} is not based on techniques similar
to those applied in the quasiconvex real valued case \cite{Volle}, nor to
those used for convex conditional maps \cite{Sca}. The idea of the proof is
to apply (\ref{111}) to the real valued quasiconvex map $\pi
_{A}:L_{t}\rightarrow \overline{\mathbb{R}}$ defined by $\pi
_{A}(X):=ess\sup_{\omega \in A}\pi (X)(\omega)$, $A\in \mathcal{F}_{s}$, and
to approximate $\pi (X)$ with 
\begin{equation*}
\pi ^{\Gamma }(X):=\sum_{A\in \Gamma }\pi _{A}(X)\mathbf{1}_{A},
\end{equation*}%
where $\Gamma $ is a finite partition of $\Omega $ of $\mathcal{F}_{s}$\
measurable sets $A\in \Gamma $. As explained in Section \ref{Sec421}, some
delicate issues arise when one tries to apply this simple and natural idea
to prove that:%
\begin{eqnarray}
&&ess\sup_{Q\in L_{t}^{\ast }\cap \mathcal{P}}\text{ }ess\inf_{\xi \in
L_{t}}\left\{ \pi (\xi )|E_{Q}[\xi |\mathcal{F}_{s}]\geq_Q E_{Q}[X|\mathcal{F%
}_{s}]\right\}  \notag \\
&=&ess\inf_{\Gamma }\text{ }ess\sup_{Q\in L_{t}^{\ast }\cap \mathcal{P}%
}ess\inf_{\xi \in L_{t}}\left\{ \pi ^{\Gamma }(\xi )|E_{Q}[\xi |\mathcal{F}%
_{s}]\geq_Q \!E_{Q}[X|\mathcal{F}_{s}]\right\}  \label{last1}
\end{eqnarray}%
The uniform approximation result here needed is stated in the key Lemma \ref%
{maggiorazione} and the Appendix \ref{a51} is devoted to prove it.

\bigskip

In this paper we limit ourselves to consider conditional maps $\pi
:L_{t}\rightarrow L_{s}$ and we defer to a forthcoming paper the study of
the temporal consistency of the family of maps $(\pi _{s})_{s\in \lbrack
0,t]}$, $\pi _{s}:L_{t}\rightarrow L_{s}$.

\bigskip

As a further motivation for our findings, we give two examples of
quasiconvex conditional maps arising in economics and finance, which will
also be analyzed in details in a forthcoming paper.

\bigskip


\begin{enumerate}
\item \textit{Certainty Equivalent in dynamic settings} . Consider a
stochastic dynamic utility (SDU) 
\begin{equation*}
u:\mathbb{R\times }[0,\infty )\times \Omega \rightarrow \mathbb{R}
\end{equation*}%
that satisfies the following conditions: the function $x\rightarrow
u(x,t,\omega )$ is strictly increasing and concave on $\mathbb{R}$, for
almost any $\omega \in \Omega $ and for $t\in \lbrack 0,\infty )$, and $%
u(x,t,\cdot )\in L^{\infty }(\Omega ,\mathcal{F}_{t},\mathbb{P})$ for all $%
(x,t)\in \mathbb{R\times }[0,\infty ).$ This functions have been recently
considered in \cite{MuZa06} and \cite{MuSoZa} to develop the theory of
forward utility.

In \cite{FM09} we study the \textit{Conditional Certainty Equivalent}\emph{\ 
}(CCE)\emph{\ }of a random variable $X\in L_{_{t}}$, which is defined as the
random variable $\pi (X)\in L_{_{s}}$ solution of the equation: 
\begin{equation*}
u(\pi (X),s)=E_{\mathbb{P} }\left[ u(X,t)|\mathcal{F}_{s}\right] .
\end{equation*}%
Thus the CCE defines the $\emph{valuation}$ operator 
\begin{equation*}
\pi :L_{t}\rightarrow L_{s},\text{ }\pi (X)=u^{-1}\left( E_{\mathbb{P} }%
\left[ u(X,t)|\mathcal{F}_{s}\right] \right) ,s).
\end{equation*}

We showed in \cite{FM09} that the CCE, as a map $\pi :L^{\infty }(\Omega ,%
\mathcal{F}_{t},\mathbb{P})\rightarrow L^{\infty }(\Omega ,\mathcal{F}_{s},%
\mathbb{P}),$ is monotone, quasi concave, regular and that for every $X\in
L^{\infty }(\Omega ,\mathcal{F}_{t},\mathbb{P})$%
\begin{equation}
\pi (X)=\inf_{Q\in \mathcal{P}}\sup_{\xi \in L^{\infty }(\mathcal{F}%
_{t})}\left\{ \pi (\xi )\mid E_{Q}[\xi |\mathcal{F}_{s}]=_Q E_{Q}[X|\mathcal{%
F}_{s}]\right\} .  \label{rapprCCE}
\end{equation}

\item \textit{Risk measures}.

\begin{enumerate}
\item \textit{Real valued quasiconvex risk measures}. Our interest in
quasiconvex analysis was triggered by the recent paper \cite{CMMMa} on
quasiconvex risk measures, where the authors shows that it is reasonable to
weaken the convexity axiom in the theory of convex risk measures, introduced
in \cite{FoSchA} and \cite{FrM}. This allows to maintain a good control of
the risk, if one also replaces cash additivity by cash subadditivity \cite%
{ER09}.

\item \textit{Dynamic risk measures}. As already mentioned the dual
representation of a conditional \emph{convex} risk measure can be found in 
\cite{Sca} and \cite{FoPe}. The findings of the present paper can be adapted
to prove the dual representation of conditional \emph{quasiconvex} risk
measures.
\end{enumerate}
\end{enumerate}

The paper is organized as follows. In Section \ref{dual} we introduce the
key definitions in order to have all the ingredients to state, in Section %
\ref{teo}, our main results. Section \ref{preliminary} is a collection of 
\emph{a priori} properties about the maps we use to obtain the dual
representation. Theorem \ref{main}, is proved in Section \ref{proof} and a
brief outline of the proof is there reported to facilitate its
understanding. The technical important Lemmas are left to the Appendix,
where we also report the proof of Theorem \ref{Volle1}.

\section{The dual representation}

\label{dual}

The probability space $(\Omega ,\mathcal{F},\mathbb{P})$ is fixed throughout
the paper and $\mathcal{G\subseteq F}$ is any sigma algebra contained in $%
\mathcal{F}$. As usual we denote with $L^{0}(\Omega ,\mathcal{F},\mathbb{P})$
the space of $\mathcal{F}$ measurable random variables that are $\mathbb{P}$
a.s. finite.

The $L^{p}(\Omega ,\mathcal{F},\mathbb{P})$ spaces, $p\in \lbrack 0,\infty
], $ will simply be denoted by $L^{p}$, unless it is necessary to specify
the sigma algebra, in which case we write $L_{\mathcal{F}}^{p}$. In presence
of an arbitrary measure $\mu $, if confusion may arise, we will explicitly
write $=_{\mu }$ (resp. $\geq _{\mu }$), meaning $\mu $ almost everywhere.
Otherwise, all equalities/inequalities among random variables are meant to
hold $\mathbb{P}$-a.s. Moreover the essential ($\mathbb{P}$ almost surely) 
\emph{supremum} $ess\sup_{\lambda }(X_{\lambda })$ of an arbitrary family of
random variables $X_{\lambda }\in L^{0}(\Omega ,\mathcal{F},\mathbb{P})$
will be simply denoted by $\sup_{\lambda }(X_{\lambda })$, and similarly for
the essential \emph{infimum} (see \cite{FoSch} Section A.5 for reference).
Here we only notice that $1_{A}\sup_{\lambda }(X_{\lambda })=\sup_{\lambda
}(1_{A}X_{\lambda })$ for any $\mathcal{F}$ measurable set $A$. Hereafter
the symbol $\hookrightarrow $ denotes inclusion and lattice embedding
between two lattices; $\vee $ (resp. $\wedge $) denotes the essential ($%
\mathbb{P}$ almost surely) \emph{maximum} (resp. the essential \emph{minimum}%
) between two random variables, which are the usual lattice operations.

We consider a lattice $L_{\mathcal{F}}:=L(\Omega ,\mathcal{F},\mathbb{P}%
)\subseteq L^{0}(\Omega ,\mathcal{F},\mathbb{P})$ and a lattice $L_{\mathcal{%
G}}:=L(\Omega ,\mathcal{G},\mathbb{P})\subseteq L^{0}(\Omega ,\mathcal{G},%
\mathbb{P})$ of $\mathcal{F}$ (resp. $\mathcal{G}$) measurable random
variables.

\begin{definition}
A map $\pi :L_{\mathcal{F}}\rightarrow L_{\mathcal{G}}$ is said to be

\begin{description}
\item[(MON)] monotone increasing if for every $X,Y\in L_{\mathcal{F}}$ 
\begin{equation*}
X\leq Y\quad \Rightarrow \quad \pi (X)\leq\pi (Y)\text{ ;}
\end{equation*}

\item[(QCO)] quasiconvex if for every $X,Y\in L_{\mathcal{F}}$, $\Lambda \in
L_{\mathcal{G}}^{0}$ and $0\leq\Lambda \leq 1$ 
\begin{equation*}
\pi (\Lambda X+(1-\Lambda )Y)\leq \pi (X)\vee \pi (Y)\text{ ;}
\end{equation*}

\item[(LSC)] $\tau -$lower semicontinuous if the set $\{X\in L_{\mathcal{F}%
}\mid \pi (X)\leq Y\}$ is closed for every $Y\in L_{\mathcal{G}}$ with
respect to a topology $\tau $ on $L_{\mathcal{F}}$.
\end{description}
\end{definition}

\begin{remark}
\label{grothe}As it happens for real valued maps, it is easy to check that
the definition of (QCO) is equivalent to the fact that all the lower level
sets 
\begin{equation*}
\mathcal{A}(Y)=\{X\in L_{\mathcal{F}}\mid \pi (X)\leq Y\}\quad \forall
\,Y\in L_{\mathcal{G}}
\end{equation*}%
are conditionally convex i.e. for all $X_{1},X_{2}\in \mathcal{A}(Y)$ and
for all $\mathcal{G}$-measurable r.v. $\Lambda $, $0\leq \Lambda \leq 1$ one
has that $\Lambda X_{1}+(1-\Lambda )X_{2}\in \mathcal{A}(Y)$.
\end{remark}

\begin{definition}
A vector space $L_{\mathcal{F}}\subseteq L_{\mathcal{F}}^{0}$ satisfies the
property $1_{\mathcal{F}}$ if 
\begin{equation}
X\in L_{\mathcal{F}}\text{ and }A\in \mathcal{F}\Longrightarrow (X\mathbf{1}%
_{A})\in L_{\mathcal{F}}\text{.}  \tag{$1_{\mathcal{F}}$}  \label{EE}
\end{equation}%
Suppose that $L_{\mathcal{F}}$ (resp. $L_{\mathcal{G}}$) satisfies the
property $(1_{\mathcal{F}})$ (resp $1_{\mathcal{G}}$). \newline
A map $\pi :L_{\mathcal{F}}\rightarrow L_{\mathcal{G}}$ is said to be

\begin{description}
\item[(REG)] regular if for every $X,Y\in L_{\mathcal{F}}$ and $A\in 
\mathcal{G}$ 
\begin{equation*}
\pi (X\mathbf{1}_{A}+Y\mathbf{1}_{A^{C}})=\pi (X)\mathbf{1}_{A}+\pi (Y)%
\mathbf{1}_{A^{C}}\text{.}
\end{equation*}
\end{description}
\end{definition}

\begin{remark}
The assumption (REG) is actually weaker than the assumption 
\begin{equation}
\pi (X\mathbf{1}_{A})=\pi (X)\mathbf{1}_{A}\quad \forall \,A\in \mathcal{G}.
\label{regStrong}
\end{equation}%
As shown in \cite{Sca}, (\ref{regStrong}) always implies (REG), and they are
equivalent if and only if $\pi (0)=0$.
\end{remark}

\subsection{The representation theorem and its consequences}

\label{teo}

\textbf{Standing assumptions}

\textit{In the sequel of the paper it is assumed that:}

\begin{description}
\item[(a)] \textit{\ }$\mathcal{G\subseteq F}$\textit{\ and the lattice }$L_{%
\mathcal{F}}$\textit{\ (resp. }$L_{\mathcal{G}}$) \textit{satisfies the
property} (1$_{\mathcal{F}}$) (resp $1_{\mathcal{G}}$)\textit{.}

\item[(b)] \textit{The order continuous dual of }$(L_{\mathcal{F}},\geq ),$ 
\textit{denoted by }$L_{\mathcal{F}}^{\ast }=(L_{\mathcal{F}},\geq )^{\ast }$%
\textit{, is a lattice ( \cite{Ali}, Th. 8.28 Ogasawara) that satisfies }$L_{%
\mathcal{F}}^{\ast }\hookrightarrow L_{\mathcal{F}}^{1}$\textit{\ and
property }(1$_{\mathcal{F}}$).

\item[(c)] \textit{The space }$L_{\mathcal{F}}$\textit{\ endowed with the
weak topology }$\sigma (L_{\mathcal{F}},L_{\mathcal{F}}^{\ast })$\textit{\
is a locally convex Riesz space. }
\end{description}

The condition (c) requires that the order continuous dual $L_{\mathcal{F}%
}^{\ast }$ is rich enough to separate the points of $L_{\mathcal{F}}$, so
that $(L_{\mathcal{F}}$\textit{,}$\sigma (L_{\mathcal{F}},L_{\mathcal{F}%
}^{\ast })$) becomes a locally convex TVS and Proposition \ref{rapprR} can
be applied.

\begin{remark}
Many important classes of spaces satisfy these conditions, as for example

\noindent - The $L^{p}$-spaces, $p\in \lbrack 1,\infty ]$: $L_{\mathcal{F}%
}=L_{\mathcal{F}}^{p},$ $L_{\mathcal{F}}^{\ast }=L_{\mathcal{F}%
}^{q}\hookrightarrow L_{\mathcal{F}}^{1}.$

\noindent - The Orlicz spaces $L^{\Psi }$ for any Young function $\Psi $: $%
L_{\mathcal{F}}=L_{\mathcal{F}}^{\Psi },$ $L_{\mathcal{F}}^{\ast }=L_{%
\mathcal{F}}^{\Psi ^{\ast }}\hookrightarrow L_{\mathcal{F}}^{1},$ where $%
\Psi ^{\ast }$ denotes the conjugate function of $\Psi $;

\noindent - The Morse subspace $M^{\Psi }$ of the Orlicz space $L^{\Psi }$,
for any continuous Young function $\Psi $: $L_{\mathcal{F}}=M_{\mathcal{F}%
}^{\Psi },$ $L_{\mathcal{F}}^{\ast }=L_{\mathcal{F}}^{\Psi ^{\ast
}}\hookrightarrow L_{\mathcal{F}}^{1}$.
\end{remark}

Set 
\begin{equation*}
\mathcal{P}=:\left\{ \frac{dQ}{d\mathbb{P}}\mid Q<<\mathbb{P}\text{ and }Q%
\text{ probability}\right\} =\left\{ \xi ^{\prime }\in L_{+}^{1}\mid E_{%
\mathbb{P} }[\xi ^{\prime }]=1\right\}
\end{equation*}%
From now on we will write with a slight abuse of notation $Q\in L_{\mathcal{F%
}}^{\ast }\cap \mathcal{P}$ instead of $\frac{dQ}{d\mathbb{P}}\in L_{%
\mathcal{F}}^{\ast }\cap \mathcal{P}$. Define for $X\in L_{\mathcal{F}}$ and 
$Q\in L_{\mathcal{F}}^{\ast }\cap \mathcal{P}$ 
\begin{equation*}
K(X,Q):=\inf_{\xi \in L_{\mathcal{F}}}\left\{ \pi (\xi )\mid E_{Q}[\xi |%
\mathcal{G}]\geq _{Q}E_{Q}[X|\mathcal{G}]\right\}
\end{equation*}%
and notice that $K(X,Q)$ depends on $X$ only through $E_{Q}[X|\mathcal{G}]$.

\begin{remark}
Since the order continuous functional on $L_{\mathcal{F}}$ are contained in $%
L^{1}$, then $Q(\xi ):=E_{Q}[\xi ]$ is well defined and finite for every $%
\xi \in L_{\mathcal{F}}$ and $Q\in L_{\mathcal{F}}^{\ast }\cap \mathcal{P}$.
In particular this and $(1_{\mathcal{F}})$ implies that $E_{Q}[\xi |\mathcal{%
G}]$ is well defined. Moreover, since $L_{\mathcal{F}}^{\ast
}\hookrightarrow L_{\mathcal{F}}^{1}$\textit{\ satisfies property }(1$_{%
\mathcal{F}}$) then $\frac{dQ}{d\mathbb{P}}1_{A}\in L_{\mathcal{F}}^{\ast }$
whenever $Q\in L_{\mathcal{F}}^{\ast }$ and $A\in \mathcal{F}.$
\end{remark}

\begin{theorem}
\label{main}If $\pi :L_{\mathcal{F}}\rightarrow L_{\mathcal{G}}$ is (MON),
(QCO), (REG) and $\sigma (L_{\mathcal{F}},L_{\mathcal{F}}^{\ast })$-LSC then 
\begin{equation}
\pi (X)=\sup_{Q\in L_{\mathcal{F}}^{\ast }\cap \mathcal{P}}K(X,Q).
\label{rapprP}
\end{equation}
\end{theorem}

Notice that in (\ref{rapprP}) the \emph{supremum} is taken over the set $L_{%
\mathcal{F}}^{\ast }\cap \mathcal{P}$. In the following corollary, proved in
Section \ref{422}, we show that we can match the conditional convex dual
representation, restricting our optimization problem over the set 
\begin{equation*}
\mathcal{P}_{\mathcal{G}}=:\left\{ \frac{dQ}{d\mathbb{P}}\mid Q\in \mathcal{P%
}\text{ and }Q=\mathbb{P}\text{ on }\mathcal{G}\right\} .
\end{equation*}%
Clearly, when $Q\in \mathcal{P}_{\mathcal{G}}$ then $L^{0}(\Omega ,\mathcal{G%
},\mathbb{P})=L^{0}(\Omega ,\mathcal{G},Q)$ and comparison of $\mathcal{G}$
measurable random variables is understood to hold indifferently for $\mathbb{%
P}$ or $Q$.

\begin{corollary}
\label{Cor1}Under the same hypothesis of Theorem \ref{main}, suppose that
for $X\in L_{\mathcal{F}}$ there exists $\eta \in L_{\mathcal{F}}$ and $%
\delta >0$ such that $\mathbb{P}(\pi (\eta )+\delta <\pi (X))=1$. Then 
\begin{equation*}
\pi (X)=\sup_{Q\in L_{\mathcal{F}}^{\ast }\cap \mathcal{P}_{\mathcal{G}%
}}K(X,Q)\text{.}
\end{equation*}
\end{corollary}

\begin{remark}
It's worth to be observed that actually the assumption (MON) is \emph{only}
used for obtaining the dual representation (\ref{rapprP}) over the set of 
\emph{positive} elements of the dual space, i.e. on probability measures
(see Proposition \ref{rapprR}). On the other hand, for $\xi ^{\prime }\in L_{%
\mathcal{F}}^{\ast }\cap L_{\mathcal{F}}^{1}$, we could define a generalized
conditional expected value $E_{\mu }[X|\mathcal{G}]=_{\mu }E_{\mathbb{P}%
}[\xi ^{\prime }X|\mathcal{G}]\cdot E_{\mathbb{P}}[\xi ^{\prime }|\mathcal{G}%
]^{-1}$, where $\mu $ is a finite \emph{signed} measure whose density is $%
\frac{d\mu }{d\mathbb{P}}=\xi ^{\prime }$ and drop the (MON) assumption in
Theorem \ref{main}. Only in the next three results the (MON)\ plays a role.
\end{remark}

\begin{definition}
We say that $\pi :L_{\mathcal{F}}\rightarrow L_{\mathcal{G}}$ is
\end{definition}

\begin{description}
\item[(CFB)] \textit{continuous from below if} 
\begin{equation*}
X_{n}\uparrow X\quad \mathbb{P}\text{ \textit{a.s.}}\quad \Rightarrow \quad
\pi (X_{n})\uparrow \pi (X)\quad \mathbb{P}\text{ \textit{a.s}.}
\end{equation*}
\end{description}

In \cite{BF09} it is proved the equivalence between: (CFB), order lsc and $%
\sigma (L_{\mathcal{F}},L_{\mathcal{F}}^{\ast })$-(LSC), for monotone convex
real valued functions. In the next proposition we show that this equivalence
remains true for monotone quasiconvex conditional maps, under the same
assumption on the topology $\sigma (L_{\mathcal{F}},L_{\mathcal{F}}^{\ast })$
adopted in \cite{BF09}.

\begin{definition}[\protect\cite{BF09}]
A linear topology $\tau $ on a Riesz space has the C-property if $X_{\alpha }%
\overset{\tau }{\rightarrow }X$ implies the existence of of a sequence $%
\{X_{\alpha _{n}}\}_{n}$ and a convex combination $Z_{n}\in conv(X_{\alpha
_{n}},...)$ such that $Z_{n}\overset{o}{\rightarrow }X$.
\end{definition}

As explained in \cite{BF09}, the assumption that $\sigma (L_{\mathcal{F}},L_{%
\mathcal{F}}^{\ast })$ has the C-property is very weak and is satisfied in
all cases of interest. When this is the case, in Theorem \ref{main} the $%
\sigma (L_{\mathcal{F}},L_{\mathcal{F}}^{\ast })$-(LSC) condition can be
replaced by (CFB), which is often easy to check.

\begin{proposition}
\label{LemmaCFB} Suppose that $\sigma (L_{\mathcal{F}},L_{\mathcal{F}}^{\ast
})$ satisfies the C-property and that $L_{\mathcal{F}}$ is order complete.
Given $\pi :L_{\mathcal{F}}\rightarrow L_{\mathcal{G}}$ satisfying (MON) and
(QCO) we have: \newline
(i) $\pi $ is $\sigma (L_{\mathcal{F}},L_{\mathcal{F}}^{\ast })$-(LSC) if
and only if (ii) $\pi $ is (CFB).
\end{proposition}

\begin{proof}
Recall that a sequence $\{X_{n}\}\subseteq L_{\mathcal{F}}$ order converge
to $X\in L_{\mathcal{F}}$, $X_{n}\overset{o}{\rightarrow }X$, if there
exists a sequence $\{Y_{n}\}\subseteq L_{\mathcal{F}}$ satisfying $%
Y_{n}\downarrow 0$ and $|X-X_{n}|\leq Y_{n}$.

(i)$\Rightarrow $ (ii): Consider $X_{n}\uparrow X$. Since $X_{n}\uparrow X$
implies $X_{n}\overset{o}{\rightarrow }X$, then for every order continuous $%
Z\in L_{\mathcal{F}}^{\ast }$ the convergence $Z(X_{n})\rightarrow Z(X)$
holds. From $L_{\mathcal{F}}^{\ast }\hookrightarrow L_{\mathcal{F}}^{1}$ 
\begin{equation*}
E_{\mathbb{P}}[ZX_{n}]\rightarrow E_{\mathbb{P}}[ZX]\quad \forall Z\in L_{%
\mathcal{F}}^{\ast }
\end{equation*}%
and we deduce that $X_{n}\overset{\sigma (L_{\mathcal{F}},L_{\mathcal{F}%
}^{\ast })}{\longrightarrow }X$. \newline
(MON) implies $\pi (X_{n})\uparrow $ and $p:=\lim_{n}\pi (X_{n})\leq \pi (X)$%
. The lower level set $\mathcal{A}_{p}=\{\xi \in L_{\mathcal{F}}\mid \pi
(\xi )\leq p\}$ is $\sigma (L_{\mathcal{F}},L_{\mathcal{F}}^{\ast })$ closed
and then $X\in \mathcal{A}_{p},$ i.e. $\pi (X)=p$.

(ii)$\Rightarrow $(i): First we prove that if $X_{n}\overset{o}{\rightarrow }%
X$ then $\pi (X)\leq \liminf_{n}\pi (X_{n})$. Define $Z_{n}:=(\inf_{k\geq
n}X_{k})\wedge X$ and note that $X-Y_{n}\leq X_{n}\leq X+Y_{n}$ implies 
\begin{equation*}
X\geq Z_{n}=\left( \inf_{k\geq n}X_{k}\right) \wedge X\geq \left(
\inf_{k\geq n}(-Y_{k})+X\right) \wedge X\uparrow X
\end{equation*}%
i.e. $Z_{n}\uparrow X$. We actually have from (MON) $Z_{n}\leq X_{n}$
implies $\pi (Z_{n})\leq \pi (X_{n})$ and from (CFB) $\pi (X)=\lim_{n}\pi
(Z_{n})\leq \liminf_{n}\pi (X_{n})$ which was our first claim. \newline
For $Y\in L_{\mathcal{G}}$ consider $\mathcal{A}_{Y}=\{\xi \in L_{\mathcal{F}%
}\mid \pi (\xi )\leq Y\}$ and a net $\{X_{\alpha }\}\subseteq L_{\mathcal{F}%
} $ such that $X_{\alpha }\overset{\sigma (L_{\mathcal{F}},L_{\mathcal{F}%
}^{\ast })}{\longrightarrow }X\in L_{\mathcal{F}}$. Since $L_{\mathcal{F}}$
satisfies the C-property, there exists $Y_{n}\in Conv(X_{\alpha _{n},...})$
such $Y_{n}\overset{o}{\rightarrow }X$. The property (QCO) implies that $%
\mathcal{A}_{Y}$ is convex and then $\{Y_{n}\}\subseteq \mathcal{A}_{Y}$.
Applying the first step we get 
\begin{equation*}
\pi (X)\leq \liminf_{n}\pi (Y_{n})\leq Y\quad \text{ i.e. }X\in \mathcal{A}%
_{Y}
\end{equation*}
\end{proof}

\bigskip

In the following Lemma and Corollary, proved in Section \ref{sec32}, we show
that the (MON) property implies that the constraint $E_{Q}[\xi |\mathcal{G}%
]\geq _{Q}E_{Q}[X|\mathcal{G}]$ may be restricted to $E_{Q}[\xi |\mathcal{G}%
]=_{Q}E_{Q}[X|\mathcal{G}]$ and that we may recover the dual representation
of a dynamic risk measure. When $Q\in L_{\mathcal{F}}^{\ast }\cap \mathcal{P}%
_{\mathcal{G}}$ the previous inequality/equality may be equivalently
intended $Q$-a.s. or $\mathbb{P}$-a.s. and so we do not need any more to
emphasize this in the notations.

\begin{lemma}
\label{mon}Suppose that for every $Q\in L_{\mathcal{F}}^{\ast }\cap \mathcal{%
P}_{\mathcal{G}}$ and $\xi \in L_{\mathcal{F}}$ we have $E_{Q}[\xi |\mathcal{%
G}]\in L_{\mathcal{F}}$. If $Q\in L_{\mathcal{F}}^{\ast }\cap \mathcal{P}_{%
\mathcal{G}}$ and if $\pi :L_{\mathcal{F}}\rightarrow L_{\mathcal{G}}$ is
(MON) and (REG) then%
\begin{equation}
K(X,Q)=\inf_{\xi \in L_{\mathcal{F}}}\left\{ \pi (\xi )\mid E_{Q}[\xi |%
\mathcal{G}]=E_{Q}[X|\mathcal{G}]\right\} .  \label{pen1}
\end{equation}
\end{lemma}

\begin{definition}
Suppose that $\pi :L_{\mathcal{F}}\rightarrow L_{\mathcal{G}}$ is convex.
The conditional Fenchel convex conjugate $\pi ^{\ast }$ of $\pi $ is given,
for $Q\in L_{\mathcal{F}}^{\ast }\cap \mathcal{P}_{\mathcal{G}}$, by the
extended valued $\mathcal{G}-$measurable random variable:%
\begin{equation*}
\pi ^{\ast }(Q)=\sup_{\xi \in L_{\mathcal{F}}}\left\{ E_{Q}[\xi |\mathcal{G}%
]-\pi (\xi )\right\} .
\end{equation*}%
A map $\pi :L_{\mathcal{F}}\rightarrow L_{\mathcal{G}}$ is said to be
\end{definition}

\begin{description}
\item[(CAS)] \textit{cash invariant if for all} $X\in L_{\mathcal{F}}$ $and$ 
$\Lambda \in L_{\mathcal{G}}$ 
\begin{equation*}
\pi (X+\Lambda )=\pi (X)+\Lambda .
\end{equation*}
\end{description}

In the literature \cite{FR1}, \cite{Sca}, \cite{FoPe} a map $\rho :L_{%
\mathcal{F}}\rightarrow L_{\mathcal{G}}$ that is monotone (decreasing),
convex, cash invariant and regular is called a \emph{convex conditional (or
dynamic) risk measure}. As a corollary of our main theorem, we deduce
immediately the dual representation of a map $\pi $ satisfying (CAS), in
terms of the Fenchel conjugate $\pi ^{\ast }$, in agreement with \cite{Sca}.
Of course, this is of no surprise since the (CAS) and (QCO) properties imply
convexity, but it supports the correctness of our dual representation.

\begin{corollary}
\label{corCAS}Suppose that for every $Q\in L_{\mathcal{F}}^{\ast }\cap 
\mathcal{P}_{\mathcal{G}}$ and $\xi \in L_{\mathcal{F}}$ we have $E_{Q}[\xi |%
\mathcal{G}]\in L_{\mathcal{F}}$.

\noindent (i) If $Q\in L_{\mathcal{F}}^{\ast }\cap \mathcal{P}_{\mathcal{G}}$
and if $\pi :L_{\mathcal{F}}\rightarrow L_{\mathcal{G}}$ is (MON), (REG) and
(CAS) then%
\begin{equation}
K(X,Q)=E_{Q}[X|\mathcal{G}]-\pi ^{\ast }(Q).  \label{KKK}
\end{equation}%
\noindent (ii) Under the same assumptions of Theorem \ref{main} and if $\pi $%
\ satisfies in addition (CAS) then 
\begin{equation*}
\pi (X)=\sup_{Q\in L_{\mathcal{F}}^{\ast }\cap \mathcal{P}_{\mathcal{G}%
}}\left\{ E_{Q}[X|\mathcal{G}]-\pi ^{\ast }(Q)\right\} \text{.}
\end{equation*}
\end{corollary}

\section{Preliminary results}

\label{preliminary}

In the sequel of the paper it is always assumed that $\pi :L_{\mathcal{F}%
}\rightarrow L_{\mathcal{G}}$ satisfies (REG).


\subsection{\label{Sec31}Properties of $R(Y,\protect\xi ^{\prime })$}

To any $\xi ^{\prime }\in L_{\mathcal{F}}^{\ast }\cap (L_{\mathcal{F}%
}^{1})_{+}$ we may associate a measure $\mu $ such that $\frac{d\mu }{d%
\mathbb{P}}=\xi ^{\prime }$. Given an arbitrary $Y\in L_{\mathcal{G}}^{0}$,
define:%
\begin{equation*}
\mathcal{A}(Y,\xi ^{\prime }):=\{\pi (\xi )\,|\,\xi \in L_{\mathcal{F}}\;E_{%
\mathbb{P}}[\xi ^{\prime }\xi |\mathcal{G}]\geq _{\mu }Y\},
\end{equation*}%
\begin{equation*}
R(Y,\xi ^{\prime }):=\inf_{\xi \in L_{\mathcal{F}}}\left\{ \pi (\xi )\mid E_{%
\mathbb{P}}[\xi ^{\prime }\xi |\mathcal{G}]\geq _{\mu }Y\right\} =\inf 
\mathcal{A}(Y,\xi ^{\prime }).
\end{equation*}

\begin{lemma}
\label{down}For every $Y\in L_{\mathcal{G}}^{0}$ and $\xi ^{\prime }\in L_{%
\mathcal{F}}^{\ast }\cap (L_{\mathcal{F}}^{1})_{+}$ the set $\mathcal{A}%
(Y,\xi ^{\prime })$ is downward directed and therefore there exists a
sequence $\left\{ \eta _{m}\right\} _{m=1}^{\infty }\in L_{\mathcal{F}}$
such that $E_{\mathbb{P}}[\xi ^{\prime }\eta _{m}|\mathcal{G}]\geq _{\mu }Y$
and as $m\uparrow \infty $, $\pi (\eta _{m})\downarrow R(Y,\xi ^{\prime })$.
\end{lemma}

\begin{proof}
We have to prove that for every $\pi (\xi _{1}),\pi (\xi _{2})\in \mathcal{A}%
(Y,\xi ^{\prime })$ there exists $\pi (\xi ^{\ast })\in \mathcal{A}(Y,\xi
^{\prime })$ such that $\pi (\xi ^{\ast })\leq \min \{\pi (\xi _{1}),\pi
(\xi _{2})\}$. Consider the $\mathcal{G}$-measurable set $G=\{\pi (\xi
_{1})\leq \pi (\xi _{2})\}$ then 
\begin{equation*}
\min \{\pi (\xi _{1}),\pi (\xi _{2})\}=\pi (\xi _{1})\mathbf{1}_{G}+\pi (\xi
_{2})\mathbf{1}_{G^{C}}=\pi (\xi _{1}\mathbf{1}_{G}+\xi _{2}\mathbf{1}%
_{G^{C}})=\pi (\xi ^{\ast }),
\end{equation*}%
where $\xi ^{\ast }=\xi _{1}\mathbf{1}_{G}+\xi _{2}\mathbf{1}_{G^{C}}$.
Since $E_{\mathbb{P}}[\xi ^{\prime }\xi ^{\ast }|\mathcal{G}]=E_{\mathbb{P}%
}[\xi ^{\prime }\xi _{1}|\mathcal{G}]\mathbf{1}_{G}+E_{\mathbb{P}}[\xi
^{\prime }\xi _{2}|\mathcal{G}]\mathbf{1}_{G^{C}}$ and $\mu<<\mathbb{P} $
together imply $E_{\mathbb{P}}[\xi ^{\prime }\xi ^{\ast }|\mathcal{G}%
]=_{\mu}E_{\mathbb{P}}[\xi ^{\prime }\xi _{1}|\mathcal{G}]\mathbf{1}_{G}+E_{%
\mathbb{P}}[\xi ^{\prime }\xi _{2}|\mathcal{G}]\mathbf{1}_{G^{C}}\geq_{\mu}
Y $, we can deduce $\pi (\xi ^{\ast })\in \mathcal{A}(Y,\xi ^{\prime })$.
\end{proof}

\begin{lemma}
\label{lemmaMinMax}Properties of $R(Y,\xi ^{\prime }).$ Let $\xi ^{\prime
}\in L_{F}^{\ast }\cap (L_{\mathcal{F}}^{1})_{+}$.

i) $R(\cdot ,\xi ^{\prime })$ is monotone

ii) $R(\lambda Y,\lambda \xi ^{\prime })=R(Y,\xi ^{\prime })$ for any $%
\lambda >0$, $Y\in L_{\mathcal{G}}$.

iii) For every $A\in \mathcal{G}$, $X\in L_{\mathcal{F}}$ and $Y=_{\mu }E_{%
\mathbb{P}}[X\xi ^{\prime }|\mathcal{G}]$ 
\begin{eqnarray}
R(Y,\xi ^{\prime })\mathbf{1}_{A} &=&\inf_{\xi \in L_{\mathcal{F}}}\left\{
\pi (\xi )\mathbf{1}_{A}\mid E_{\mathbb{P}}[\xi ^{\prime }\xi |\mathcal{G}%
]\geq _{\mu }Y\right\}  \label{TakingOut1} \\
&=&\inf_{\xi \in L_{\mathcal{F}}}\left\{ \pi (\xi )\mathbf{1}_{A}\mid E_{%
\mathbb{P}}[\xi ^{\prime }\xi \mathbf{1}_{A}|\mathcal{G}]\geq _{\mu }Y%
\mathbf{1}_{A}\right\} ,  \label{TakingOut}
\end{eqnarray}

iv) For every $Y_{1},Y_{2}\in L_{\mathcal{G}}$

\qquad (a) $R(Y_{1},\xi ^{\prime })\wedge R(Y_{2},\xi ^{\prime
})=R(Y_{1}\wedge Y_{2},\xi ^{\prime })$

\qquad (b) $R(Y_{1},\xi ^{\prime })\vee R(Y_{2},\xi ^{\prime })=R(Y_{1}\vee
Y_{2},\xi ^{\prime })$

v) The map $R(Y,\xi ^{\prime })$ is quasi-affine with respect to $Y$ in the
sense that for every $Y_{1},Y_{2},\Lambda \,\in L_{\mathcal{G}}$ and $0\leq
\Lambda \leq 1,$ we have

$\qquad R(\Lambda Y_{1}+(1-\Lambda )Y_{2},\xi ^{\prime })\geq R(Y_{1},\xi
^{\prime })\wedge R(Y_{2},\xi ^{\prime })\text{\quad (quasiconcavity)}$

$\qquad R(\Lambda Y_{1}+(1-\Lambda )Y_{2},\xi ^{\prime })\leq R(Y_{1},\xi
^{\prime })\vee R(Y_{2},\xi ^{\prime })\text{\quad (quasiconvexity).}$
\end{lemma}

\begin{proof}
Since $\pi (\xi \mathbf{1}_{A})-\pi (0)\mathbf{1}_{A^{C}}=\pi (\xi )\mathbf{1%
}_{A}$ , w.l.o.g. we may assume in the sequel of this proof that $\pi (0)=0$
and so $\pi (\xi \mathbf{1}_{A})=\pi (\xi )\mathbf{1}_{A}.$

\noindent (i) and (ii) are trivial.

\noindent (iii) By definition of the essential infimum one easily deduce (%
\ref{TakingOut1}). Let $\frac{d\mu }{d\mathbb{P}}=\xi ^{\prime } $. To prove
(\ref{TakingOut}), for every $\xi \in L_{\mathcal{F}}$ such that $E_{\mathbb{%
P}}[\xi ^{\prime }\xi \mathbf{1}_{A}|\mathcal{G}]\geq _{\mu }Y\mathbf{1}_{A}$
we define the random variable $\eta =\xi \mathbf{1}_{A}+X\mathbf{1}_{A^{C}}$
which satisfies $E_{\mathbb{P}}[\xi ^{\prime }\eta |\mathcal{G}]\geq _{\mu
}Y $. In fact since $\mu <<\mathbb{P}$ we have that $E_{\mathbb{P}}[\xi
^{\prime }\eta |\mathcal{G}]=E_{\mathbb{P}}[\xi ^{\prime }\xi |\mathcal{G}]%
\mathbf{1}_{A}+E_{\mathbb{P}}[\xi ^{\prime }X|\mathcal{G}]\mathbf{1}_{A^{C}}$
implies 
\begin{equation*}
E_{\mathbb{P}}[\xi ^{\prime }\eta |\mathcal{G}]=_{\mu }E_{\mathbb{P}}[\xi
^{\prime }\xi |\mathcal{G}]\mathbf{1}_{A}+E_{\mathbb{P}}[\xi ^{\prime }X|%
\mathcal{G}]\mathbf{1}_{A^{C}}\geq _{\mu }Y.
\end{equation*}%
Therefore 
\begin{equation*}
\left\{ \eta \mathbf{1}_{A}\mid \eta \in L_{\mathcal{F}}\text{, }E_{\mathbb{P%
}}[\xi ^{\prime }\eta |\mathcal{G}]\geq _{\mu }Y\right\} =\left\{ \xi 
\mathbf{1}_{A}\mid \xi \in L_{\mathcal{F}}\text{, }E_{\mathbb{P}}[\xi
^{\prime }\xi \mathbf{1}_{A}|\mathcal{G}]\geq _{\mu }Y\mathbf{1}_{A}\right\}
\end{equation*}%
Hence from (\ref{TakingOut1}):%
\begin{eqnarray*}
\mathbf{1}_{A}R(Y,\xi ^{\prime }) &=&\inf_{\eta \in L_{\mathcal{F}}}\left\{
\pi (\eta \mathbf{1}_{A})\mid E_{\mathbb{P}}[\xi ^{\prime }\eta |\mathcal{G}%
]\geq _{\mu }Y\right\} \\
&=&\inf_{\xi \in L_{\mathcal{F}}}\left\{ \pi (\xi \mathbf{1}_{A})\mid E_{%
\mathbb{P}}[\xi ^{\prime }\xi \mathbf{1}_{A}|\mathcal{G}]\geq _{\mu }Y%
\mathbf{1}_{A}\right\}
\end{eqnarray*}%
and (\ref{TakingOut}) follows.

\noindent iv) a): Since $R(\cdot ,\xi ^{\prime })$ is monotone, the
inequalities $R(Y_{1},\xi ^{\prime })\wedge R(Y_{2},\xi ^{\prime })\geq
R(Y_{1}\wedge Y_{2},\xi ^{\prime })$ and $R(Y_{1},\xi ^{\prime })\vee
R(Y_{2},\xi ^{\prime })\leq R(Y_{1}\vee Y_{2},\xi ^{\prime })$ are always
true.

\noindent To show the opposite inequalities, define the $\mathcal{G}$%
-measurable sets: $B:=\{R(Y_{1},\xi ^{\prime })\leq R(Y_{2},\xi ^{\prime
})\} $ and $A:=\{Y_{1}\leq Y_{2}\}$ so that 
\begin{equation}
R(Y_{1},\xi ^{\prime })\wedge R(Y_{2},\xi ^{\prime })=R(Y_{1},\xi ^{\prime })%
\mathbf{1}_{B}+R(Y_{2},\xi ^{\prime })\mathbf{1}_{B^{C}}\leq R(Y_{1},\xi
^{\prime })\mathbf{1}_{A}+R(Y_{2},\xi ^{\prime })\mathbf{1}_{A^{C}}
\label{RR}
\end{equation}%
\begin{equation*}
R(Y_{1},\xi ^{\prime })\vee R(Y_{2},\xi ^{\prime })=R(Y_{1},\xi ^{\prime })%
\mathbf{1}_{B^{C}}+R(Y_{2},\xi ^{\prime })\mathbf{1}_{B}\geq R(Y_{1},\xi
^{\prime })\mathbf{1}_{A^{C}}+R(Y_{2},\xi ^{\prime })\mathbf{1}_{A}
\end{equation*}%
Set: $D(A,Y)=\left\{ \xi \mathbf{1}_{A}\mid \xi \in L_{\mathcal{F}}\text{, }%
E_{\mathbb{P}}[\xi ^{\prime }\xi \mathbf{1}_{A}|\mathcal{G}]\geq _{\mu }Y%
\mathbf{1}_{A}\right\} $ and check that 
\begin{equation*}
D(A,Y_{1})+D(A^{C},Y_{2})=\left\{ \xi \in L_{\mathcal{F}}\mid E_{\mathbb{P}%
}[\xi ^{\prime }\xi |\mathcal{G}]\geq _{\mu }Y_{1}\mathbf{1}_{A}+Y_{2}%
\mathbf{1}_{A^{C}}\right\} :=D
\end{equation*}%
From (\ref{RR}) \ and using (\ref{TakingOut}) we get:%
\begin{eqnarray*}
R(Y_{1},\xi ^{\prime })\wedge R(Y_{2},\xi ^{\prime })\leq &&R(Y_{1},\xi
^{\prime })\mathbf{1}_{A}+R(Y_{2},\xi ^{\prime })\mathbf{1}_{A^{C}} \\
= &&\inf_{\xi \mathbf{1}_{A}\in D(A,Y_{1})}\left\{ \pi (\xi \mathbf{1}%
_{A})\right\} +\inf_{\eta \mathbf{1}_{A^{C}}\in D(A^{C},Y_{2})}\left\{ \pi
(\eta \mathbf{1}_{A^{C}})\right\} \\
= &&\inf_{\substack{ \xi \mathbf{1}_{A}\in D(A,Y_{1})  \\ \eta \mathbf{1}%
_{A^{C}}\in D(A^{C},Y_{2})}}\left\{ \pi (\xi \mathbf{1}_{A})+\pi (\eta 
\mathbf{1}_{A^{C}})\right\} \\
= &&\inf_{(\xi \mathbf{1}_{A}+\eta \mathbf{1}_{A^{C}})\in
D(A,Y_{1})+D(A^{C},Y_{2})}\left\{ \pi (\xi \mathbf{1}_{A}+\eta \mathbf{1}%
_{A^{C}})\right\} \\
= &&\inf_{\xi \in D}\left\{ \pi (\xi )\right\} = R(Y_{1}\mathbf{1}_{A}+Y_{2}%
\mathbf{1}_{A^{C}},\xi ^{\prime })= R(Y_{1}\wedge Y_{2},\xi ^{\prime }).
\end{eqnarray*}%
\emph{Simile modo: }iv) $b).$

\noindent (v) From the monotonicity of $R(\cdot ,\xi ^{\prime })$, $%
R(Y_{1}\wedge Y_{2},\xi ^{\prime })\leq R(\Lambda Y_{1}+(1-\Lambda
)Y_{2},\xi ^{\prime })$ (resp. $R(Y_{1}\vee Y_{2},\xi ^{\prime })\geq
R(\Lambda Y_{1}+(1-\Lambda )Y_{2},\xi ^{\prime })$) and then the thesis
follows from iv).
\end{proof}

\subsection{Properties of K$(X,Q)\label{sec32}$}

For $\xi ^{\prime }\in L_{\mathcal{F}}^{\ast }\cap (L_{\mathcal{F}}^{1})_{+}$
and $X\in L_{\mathcal{F}}$ 
\begin{equation*}
R(E_{\mathbb{P}}[\xi ^{\prime }X|\mathcal{G}],\xi ^{\prime })=\inf_{\xi \in
L_{\mathcal{F}}}\left\{ \pi (\xi )\mid E_{\mathbb{P}}[\xi ^{\prime }\xi |%
\mathcal{G}]\geq _{\mu }E_{\mathbb{P}}[\xi ^{\prime }X|\mathcal{G}]\right\}
=K(X,\xi ^{\prime }).
\end{equation*}%
Notice that $K(X,\xi ^{\prime })=K(X,\lambda \xi ^{\prime })\text{ for every 
}\lambda >0$ and thus we can consider $K(X,\xi ^{\prime })$, $\xi ^{\prime
}\neq 0$, always defined on the normalized elements $Q\in L_{\mathcal{F}%
}^{\ast }\cap \mathcal{P}$.

\noindent Moreover, from $E_{\mathbb{P}}\left[ \frac{dQ}{d\mathbb{P}}X|%
\mathcal{G}\right] =_{Q}E_{\mathbb{P}}[\frac{dQ}{d\mathbb{P}}\mid \mathcal{G}%
]E_{Q}[X|\mathcal{G}]$ and $E_{\mathbb{P}}[\frac{dQ}{d\mathbb{P}}\mid 
\mathcal{G}]>_{Q}0$ we deduce: 
\begin{equation*}
E_{\mathbb{P}}\left[ \frac{dQ}{d\mathbb{P}}\xi \mid \mathcal{G}\right] \geq
_{Q}E_{\mathbb{P}}\left[ \frac{dQ}{d\mathbb{P}}X\mid \mathcal{G}\right]
\Longleftrightarrow E_{Q}[\xi |\mathcal{G}]\geq _{Q}E_{Q}[X|\mathcal{G}].
\end{equation*}%
For $Q\in L_{\mathcal{F}}^{\ast }\cap \mathcal{P}$ we then set:%
\begin{equation*}
K(X,Q):=\inf_{\xi \in L_{\mathcal{F}}}\left\{ \pi (\xi )\mid E_{Q}[\xi |%
\mathcal{G}]\geq _{Q}E_{Q}[X|\mathcal{G}]\right\} =R\left( E_{\mathbb{P}}%
\left[ \frac{dQ}{d\mathbb{P} }X\mid \mathcal{G}\right] ,\frac{dQ}{d\mathbb{P}
}\right) .
\end{equation*}

\begin{lemma}
\label{upwardn}Properties of $K(X,Q)$. Let $Q\in L_{\mathcal{F}}^{\ast }\cap 
\mathcal{P}$ and $X\in L_{\mathcal{F}}.$

\noindent i) $K(\cdot ,Q)$ is monotone and quasi affine.

\noindent ii) $K(X,\cdot )$ is positively homogeneous.

\noindent iii) $K(X,Q)\mathbf{1}_{A}=\inf_{\xi \in L_{\mathcal{F}}}\left\{
\pi (\xi )\mathbf{1}_{A}\mid E_{Q}[\xi \mathbf{1}_{A}|\mathcal{G}]\geq_Q
E_{Q}[X\mathbf{1}_{A}|\mathcal{G}]\right\} $ for all $A\in \mathcal{G}.$

\noindent iv) There exists a sequence $\left\{ \xi _{m}^{Q}\right\}
_{m=1}^{\infty }\in L_{\mathcal{F}}$ such that 
\begin{equation*}
E_{Q}[\xi _{m}^{Q}|\mathcal{G}]\geq_Q E_{Q}[X|\mathcal{G}]\quad \forall
\,m\geq 1,\quad \pi (\xi _{m}^{Q})\downarrow K(X,Q)\quad \text{as }m\uparrow
\infty .
\end{equation*}%
\noindent v) The set $\mathcal{K}=\left\{ K(X,Q)\mid Q\in L_{\mathcal{F}%
}^{\ast }\cap \mathcal{P}\right\} $ is upward directed, i.e. for every $%
K(X,Q_{1}),$ $K(X,Q_{2})\in \mathcal{K}$ there exists $K(X,\widehat{Q})\in 
\mathcal{K}$ such that $K(X,\widehat{Q})\geq K(X,Q_{1})\vee K(X,Q_{2})$.

\noindent vi) Let $Q_{1}$ and $Q_{2}$ be elements of $L_{\mathcal{F}}^{\ast
}\cap \mathcal{P}$ and $B\in \mathcal{G}.$ If $\frac{dQ_{1}}{d\mathbb{P} }%
\mathbf{1}_{B}=\frac{dQ_{2}}{d\mathbb{P} }\mathbf{1}_{B}$ then $K(X,Q_{1})%
\mathbf{1}_{B}=K(X,Q_{2})\mathbf{1}_{B}.$
\end{lemma}

\begin{proof}
The monotonicity property in (i), (ii) and (iii) are trivial; from Lemma \ref%
{lemmaMinMax} v) it follows that $K(\cdot ,Q)$ is quasi affine; (iv) is an
immediate consequence of Lemma \ref{down}.

(v) Define $F=\{K(X,Q_{1})\geq K(X,Q_{2})\}$ and let $\widehat{Q}$ given by $%
\frac{d\widehat{Q}}{d\mathbb{P}}:=\mathbf{1}_{F}\frac{dQ_{1}}{d\mathbb{P}}+%
\mathbf{1}_{F^{C}}\frac{dQ_{2}}{d\mathbb{P}}$; up to a normalization factor
(from property (ii)) we may suppose $\widehat{Q}\in L_{\mathcal{F}}^{\ast
}\cap \mathcal{P}$. We need to show that 
\begin{equation*}
K(X,\widehat{Q})=K(X,Q_{1})\vee K(X,Q_{2})=K(X,Q_{1})\mathbf{1}%
_{F}+K(X,Q_{2})\mathbf{1}_{F^{C}}.
\end{equation*}%
From $E_{\widehat{Q}}[\xi |\mathcal{G}]=_{\widehat{Q}}E_{Q_{1}}[\xi |%
\mathcal{G}]\mathbf{1}_{F}+E_{Q_{2}}[\xi |\mathcal{G}]\mathbf{1}_{F^{C}}$ we
get $E_{\widehat{Q}}[\xi |\mathcal{G}]\mathbf{1}_{F}=_{Q_{1}}E_{Q_{1}}[\xi |%
\mathcal{G}]\mathbf{1}_{F}$ and $E_{\widehat{Q}}[\xi |\mathcal{G}]\mathbf{1}%
_{F^{C}}=_{Q_{2}}E_{Q_{2}}[\xi |\mathcal{G}]\mathbf{1}_{F^{C}}$. In the
second place, for $i=1,2$, consider the sets 
\begin{equation*}
\widehat{A}=\{\xi \in L_{\mathcal{F}}\mid E_{\widehat{Q}}[\xi |\mathcal{G}%
]\geq _{\widehat{Q}}E_{\widehat{Q}}[X|\mathcal{G}]\}\quad A_{i}=\{\xi \in L_{%
\mathcal{F}}\mid E_{Q_{i}}[\xi |\mathcal{G}]\geq _{Q_{i}}E_{Q_{i}}[X|%
\mathcal{G}]\}.
\end{equation*}%
For every $\xi \in A_{1}$ define $\eta =\xi \mathbf{1}_{F}+X\mathbf{1}%
_{F^{C}}$ 
\begin{eqnarray*}
Q_{1}<<\mathbb{P}\;\Rightarrow &\;\eta \mathbf{1}_{F}=_{Q_{1}}\xi \mathbf{1}%
_{F}\;\Rightarrow &\;E_{\widehat{Q}}[\eta |\mathcal{G}]\mathbf{1}_{F}\geq _{%
\widehat{Q}}E_{\widehat{Q}}[X|\mathcal{G}]\mathbf{1}_{F} \\
Q_{2}<<\mathbb{P}\;\Rightarrow &\;\eta \mathbf{1}_{F^{C}}=_{Q_{2}}X\mathbf{1}%
_{F^{C}}\;\Rightarrow &\;E_{\widehat{Q}}[\eta |\mathcal{G}]\mathbf{1}%
_{F^{C}}=_{\widehat{Q}}E_{\widehat{Q}}[X|\mathcal{G}]\mathbf{1}_{F^{C}}
\end{eqnarray*}%
Then $\eta \in \widehat{A}$ and $\pi (\xi )\mathbf{1}_{F}=\pi (\xi \mathbf{1}%
_{F})-\pi (0)\mathbf{1}_{F^{C}}=\pi (\eta \mathbf{1}_{F})-\pi (0)\mathbf{1}%
_{F^{C}}=\pi (\eta )\mathbf{1}_{F}$. \newline
Viceversa, for every $\eta \in \widehat{A}$ define $\xi =\eta \mathbf{1}%
_{F}+X\mathbf{1}_{F^{C}}$. Then $\xi \in A_{1}$ and again $\pi (\xi )\mathbf{%
1}_{F}=\pi (\eta )\mathbf{1}_{F}$. Hence 
\begin{equation*}
\inf_{\xi \in A_{1}}\pi (\xi )\mathbf{1}_{F}=\inf_{\eta \in \widehat{A}}\pi
(\eta )\mathbf{1}_{F}.
\end{equation*}%
In a similar way: $\inf_{\xi \in A_{2}}\pi (\xi )\mathbf{1}%
_{F^{C}}=\inf_{\eta \in \widehat{A}}\pi (\eta )\mathbf{1}_{F^{C}}$ and we
can finally deduce $K(X,Q_{1})\vee K(X,Q_{2})=K(X,\widehat{Q}).$

(vi). By the same argument used in (v), it can be shown that $\inf_{\xi \in
A_{1}}\pi (\xi )\mathbf{1}_{B}=\inf_{\xi \in A_{2}}\pi (\xi )\mathbf{1}_{B}$
and the thesis.
\end{proof}

\bigskip

\begin{proof}[Proof of Lemma \protect\ref{mon}]
Let us denote with $r(X,Q)$ the right hand side of equation (\ref{pen1}) and
notice that $K(X,Q)\leq r(X,Q)$. By contradiction, suppose that $\mathbb{P}%
(A)>0$ where $A=:\{K(X,Q)<r(X,Q)\}$. As shown in Lemma \ref{upwardn} iv),
there exists a r.v. $\xi \in L_{\mathcal{F}}$ satisfying the following
conditions

\begin{itemize}
\item $E_{Q}[\xi |\mathcal{G}]\geq _{Q}E_{Q}[X|\mathcal{G}]$ and $%
Q(E_{Q}[\xi |\mathcal{G}]>E_{Q}[X|\mathcal{G}])>0$.

\item $K(X,Q)(\omega )\leq \pi (\xi )(\omega )<r(X,Q)(\omega )$ for $\mathbb{%
P}$-almost every $\omega \in B\subseteq A$ and $\mathbb{P}(B)>0$.
\end{itemize}

Set $Z=_{Q}E_{Q}[\xi -X|\mathcal{G}].$ By assumption, $Z\in L_{\mathcal{F}}$
and it satisfies $Z\geq _{Q}0$ and, since $Q\in \mathcal{P}_{\mathcal{G}}$, $%
Z\geq 0$. Then, thanks to (MON), $\pi (\xi )\geq\pi (\xi -Z)$. From $%
E_{Q}[\xi -Z|\mathcal{G}]=_{Q}E_{Q}[X|\mathcal{G}]$ we deduce: 
\begin{equation*}
K(X,Q)(\omega )\leq \pi (\xi )(\omega )<r(X,Q)(\omega )\leq \pi (\xi
-Z)(\omega )\text{ for }\mathbb{P}\text{-a.e. }\omega \in B,
\end{equation*}%
which is a contradiction.
\end{proof}

\bigskip

\begin{proof}[Proof of Corollary \protect\ref{corCAS}]
The (CAS) property implies that for every $X\in L_{\mathcal{F}}$ and $\delta
>0$, $\mathbb{P}(\pi (X-2\delta )+\delta <\pi (X))=1$. So the hypothesis of
Corollary \ref{Cor1} holds true and we only need to prove (\ref{KKK}), since
(ii) is a consequence of (i) and Corollary \ref{Cor1}. Let $Q\in L_{\mathcal{%
F}}^{\ast }\cap \mathcal{P}_{\mathcal{G}}$. Applying Lemma \ref{mon} we
deduce:%
\begin{eqnarray*}
K(X,Q)= &&\inf_{\xi \in L_{\mathcal{F}}}\left\{ \pi (\xi )\mid E_{Q}[\xi |%
\mathcal{G}]=_{Q}E_{Q}[X|\mathcal{G}]\right\} \\
= &&E_{Q}[X|\mathcal{G}]+\inf_{\xi \in L_{\mathcal{F}}}\left\{ \pi (\xi
)-E_{Q}[X|\mathcal{G}]\mid E_{Q}[\xi |\mathcal{G}]=_{Q}E_{Q}[X|\mathcal{G}%
]\right\} \\
= &&E_{Q}[X|\mathcal{G}]+\inf_{\xi \in L_{\mathcal{F}}}\left\{ \pi (\xi
)-E_{Q}[\xi |\mathcal{G}]\mid E_{Q}[\xi |\mathcal{G}]=_{Q}E_{Q}[X|\mathcal{G}%
]\right\} \\
= &&E_{Q}[X|\mathcal{G}]-\sup_{\xi \in L_{\mathcal{F}}}\left\{ E_{Q}[\xi |%
\mathcal{G}]-\pi (\xi )\mid E_{Q}[\xi |\mathcal{G}]=_{Q}E_{Q}[X|\mathcal{G}%
]\right\} \\
= &&E_{Q}[X|\mathcal{G}]-\pi ^{\ast }(Q),
\end{eqnarray*}%
where the last equality follows from $Q\in \mathcal{P}_{\mathcal{G}}$ and 
\begin{eqnarray*}
\pi ^{\ast }(Q)= &&\sup_{\xi \in L_{\mathcal{F}}}\left\{ E_{Q}[\xi
+E_{Q}[X-\xi |\mathcal{G}]\mid \mathcal{G}]-\pi (\xi +E_{Q}[X-\xi |\mathcal{G%
}])\right\} \\
= &&\sup_{\eta \in L_{\mathcal{F}}}\left\{ E_{Q}[\eta |\mathcal{G}]-\pi
(\eta )\mid \eta =\xi +E_{Q}[X-\xi |\mathcal{G}]\right\} \\
\leq &&\sup_{\xi \in L_{\mathcal{F}}}\left\{ E_{Q}[\xi |\mathcal{G}]-\pi
(\xi )\mid E_{Q}[\xi |\mathcal{G}]=_{Q}E_{Q}[X|\mathcal{G}]\right\} \leq \pi
^{\ast }(Q).
\end{eqnarray*}
\end{proof}

\subsection{On H(X) and a first approximation\label{s33}}

\label{approximation}

For $X\in L_{\mathcal{F}}$ we set 
\begin{equation*}
H(X):=\sup_{Q\in L_{\mathcal{F}}^{\ast }\cap \mathcal{P}}K(X,Q)=\sup_{Q\in
L_{\mathcal{F}}^{\ast }\cap \mathcal{P}}\inf_{\xi \in L_{\mathcal{F}%
}}\left\{ \pi (\xi )\mid E_{Q}[\xi |\mathcal{G}]\geq_Q E_{Q}[X|\mathcal{G}%
]\right\}
\end{equation*}%
and notice that for all $A\in \mathcal{G}$ 
\begin{equation*}
H(X)\mathbf{1}_{A}=\sup_{Q\in L_{\mathcal{F}}^{\ast }\cap \mathcal{P}%
}\inf_{\xi \in L_{\mathcal{F}}}\left\{ \pi (\xi )\mathbf{1}_{A}\mid
E_{Q}[\xi |\mathcal{G}]\geq_Q E_{Q}[X|\mathcal{G}]\right\} .
\end{equation*}

\begin{lemma}
\label{LH}Properties of $H(X)$. Let $X\in L_{\mathcal{F}}$.

i) $H$ is monotone

ii) $H(X\mathbf{1}_{A})\mathbf{1}_{A}=H(X)\mathbf{1}_{A}$ for any $A\in 
\mathcal{G}$ $.$

iii) There exist a sequence $\left\{ Q^{k}\right\} _{k\geq 1}\in L_{\mathcal{%
F}}^{\ast }$ and, for each $k\geq 1$,\ a sequence $\left\{ \xi
_{m}^{Q^{k}}\right\} _{m\geq 1}\in L_{\mathcal{F}}$ satisfying $%
E_{Q^{k}}[\xi _{m}^{Q^{k}}\mid \mathcal{G}]\geq _{Q^{k}}E_{Q^{k}}[X|\mathcal{%
G}]$ and 
\begin{eqnarray}
\pi (\xi _{m}^{Q^{k}}) &\downarrow &K(X,Q^{k})\text{ as }m\uparrow \infty 
\text{, }K(X,Q^{k})\uparrow H(X)\text{ as }k\uparrow \infty ,  \label{lim2}
\\
H(X) &=&\lim_{k\rightarrow \infty }\lim_{m\rightarrow \infty }\pi (\xi
_{m}^{Q^{k}}).  \label{lim}
\end{eqnarray}
\end{lemma}

\begin{proof}
i) is trivial; ii) follows applying the same argument used in equation (\ref%
{TakingOut}); the other property is an immediate consequence of what proved
in Lemma \ref{upwardn} and \ref{down} regarding the properties of being
downward directed and upward directed.
\end{proof}

\begin{lemma}
Let $\mathcal{Q}\subseteq L_{F}^{\ast }\cap \mathcal{P}$ and suppose that
the map $S:$ $L_{\mathcal{G}}\times \mathcal{Q}\rightarrow L_{\mathcal{G}}$
is quasiconvex with respect to $Y\in L_{\mathcal{G}}$, for each $Q\in 
\mathcal{Q}.$ Then the functional 
\begin{equation*}
f(X)=\sup_{Q\in \mathcal{Q}}S(E_{Q}[X|\mathcal{G}],Q)
\end{equation*}%
is quasiconvex with respect to $X\in L_{\mathcal{F}}$. In particular, $H(X)$
is quasiconvex with respect to $X\in L_{\mathcal{F}}$.
\end{lemma}

\begin{proof}
The first claim is a straightforward application of the definition. By Lemma %
\ref{upwardn} i) $K(\cdot ,Q)$ is quasiconvex and the second statement
follows.
\end{proof}

\bigskip

The following Proposition is an uniform approximation result which stands
under stronger assumptions, that are satisfied, for example, by $L^{p}$
spaces, $p\in \lbrack 1,+\infty ]$. We will not use this Proposition in the
proof of Theorem \ref{main}, even though it can be useful for understanding
the heuristic outline of its proof, as sketched in Section \ref{Sec421}.

\begin{proposition}
\label{LQQ}Suppose that $L_{F}^{\ast }\hookrightarrow L_{\mathcal{F}}^{1}$
is a Banach Lattice with the property: for any sequence $\{\eta
_{n}\}_{n}\subseteq (L_{F}^{\ast })_{+}$, $\eta _{n}\eta _{m}=0$ for every $%
n\neq m$, there exists a sequence $\{\alpha _{k}\}_{k}\subset (0,+\infty )$
such that $\sum_{n}\alpha _{n}\eta _{n}\in (L_{F}^{\ast })_{+}$. If $%
H(X)>-\infty $ $\mathbb{P}-a.s.,$ then for every $\varepsilon >0$ there
exists $Q_{\varepsilon }\in L_{F}^{\ast }\cap \mathcal{P}$ such that 
\begin{equation}
H(X)-K(X,Q_{\varepsilon })<\varepsilon  \label{appr}
\end{equation}
\end{proposition}

\begin{proof}
From Lemma \ref{LH}, eq. (\ref{lim2}), we know that there exists a sequence $%
Q_{k}\in L_{\mathcal{F}}^{\ast }\cap \mathcal{P}$ such that:%
\begin{equation*}
K(X,Q_{k})\uparrow H(X)\text{, as }k\uparrow \infty .
\end{equation*}%
Define for each $k\geq 1$ the sets 
\begin{equation*}
D_{k}=:\left\{ H(X)-K(X,Q_{k})\leq \varepsilon \right\}
\end{equation*}%
and note that 
\begin{equation}
\mathbb{P}\left( D_{k}\right) \uparrow 1\text{ as }k\uparrow \infty .
\label{DD}
\end{equation}%
Consider the disjoint family $\left\{ F_{k}\right\} _{k\geq 1}$ of $\mathcal{%
G}-$measurable sets: $F_{1}=D_{1},\quad F_{k}=D_{k}\setminus D_{k-1}$, $%
k\geq 2.$ By induction one easily shows that $\bigcup%
\limits_{k=1}^{n}F_{k}=D_{n}$ for all $n\geq 1.$ This and (\ref{DD}) imply
that $\mathbb{P}\left( \bigcup\limits_{k=1}^{\infty }F_{k}\right) =1$.
Consider the sequence $\left\{ \frac{dQ_{k}}{d\mathbb{P}}\mathbf{1}%
_{F_{k}}\right\} $. From the assumption on $L_{\mathcal{F}}^{\ast }$ we may
find a sequence $\{\alpha _{k}\}_{k}\subset (0,+\infty )$ such that $\frac{d%
\widetilde{Q}_{\epsilon }}{d\mathbb{P}}=:\sum_{k=1}^{\infty }\alpha _{k}%
\frac{dQ_{k}}{d\mathbb{P}}\mathbf{1}_{F_{k}}\in L_{\mathcal{F}}^{\ast
}\hookrightarrow L_{\mathcal{F}}^{1}$. Hence, $\widetilde{Q}_{\epsilon }\in
(L_{\mathcal{F}}^{\ast })_{+}\cap (L_{\mathcal{F}}^{1})_{+}$ and, since $%
\left\{ F_{k}\right\} _{k\geq 1}$ are disjoint, 
\begin{equation*}
\frac{d\widetilde{Q}_{\epsilon }}{d\mathbb{P}}\mathbf{1}_{F_{k}}=\alpha _{k}%
\frac{dQ_{k}}{d\mathbb{P}}\mathbf{1}_{F_{k}},\text{ for any }k\geq 1.
\end{equation*}%
Normalize $\widetilde{Q}_{\epsilon }$ and denote with $Q_{\varepsilon
}=\lambda \widetilde{Q}_{\epsilon }\in L_{\mathcal{F}}^{\ast }\cap \mathcal{P%
}$ the element satisfying $\parallel \frac{dQ_{\varepsilon }}{d\mathbb{P}}%
\parallel _{L_{F}^{1}}=1.$ Applying Lemma \ref{upwardn} (vi) we deduce that
for any $k\geq 1$ 
\begin{equation*}
K(X,Q_{\epsilon })\mathbf{1}_{F_{k}}=K(X,\widetilde{Q}_{\epsilon })\mathbf{1}%
_{F_{k}}=K(X,\alpha _{k}Q_{k})\mathbf{1}_{F_{k}}=K(X,Q_{k})\mathbf{1}%
_{F_{k}},
\end{equation*}%
and 
\begin{equation*}
H(X)\mathbf{1}_{F_{k}}-K(X,Q_{\varepsilon })\mathbf{1}_{F_{k}}=H(X)\mathbf{1}%
_{F_{k}}-K(X,Q_{k})\mathbf{1}_{F_{k}}\leq \varepsilon \mathbf{1}_{F_{k}}.
\end{equation*}%
The condition (\ref{appr}) is then a consequence of $\mathbb{P}\left(
D_{k}\right) \uparrow 1$. Notice that the assumption $H(X)>-\infty $ is only
used in (\ref{DD}) and that without this assumption the conclusion (\ref%
{appr}) would hold true on the set $\left\{ H(X)>-\infty \right\} .$
\end{proof}

\subsection{On the map $\protect\pi _{A}$}

Consider the following

\begin{definition}
Given $\pi :L_{\mathcal{F}}\rightarrow L_{\mathcal{G}}$ we define for every $%
A\in \mathcal{G}$ the map 
\begin{equation*}
\pi _{A}:L_{\mathcal{F}}\rightarrow \overline{\mathbb{R}}\;\text{ by }\;\pi
_{A}(X):=ess\sup_{\omega \in A}\pi (X)(\omega).
\end{equation*}
\end{definition}

Notice that the map $\pi _{A}$ inherits from $\pi $ the properties (MON),
(QCO) and (CFB). Applying Proposition \ref{LemmaCFB} we deduce that $\pi
_{A} $ is also $\sigma (L_{\mathcal{F}},L_{\mathcal{F}}^{\ast })$-lsc.

\begin{proposition}
\label{rapprPiA}Under the same assumptions of Theorem \ref{main} if $A\in 
\mathcal{G}$ 
\begin{equation}
\pi _{A}(X)=\sup_{Q\in L_{\mathcal{F}}^{\ast }\cap \mathcal{P}}\inf_{\xi \in
L_{\mathcal{F}}}\left\{ \pi _{A}(\xi )\mid E_{Q}[\xi |\mathcal{G}]\geq_Q
E_{Q}[X|\mathcal{G}]\right\} .  \label{piAsup}
\end{equation}
\end{proposition}

\begin{proof}
From $L_{\mathcal{F}}^{\ast }\hookrightarrow L^{1}(\mathcal{F}),$ we have: $%
L_{\mathcal{F}}^{\ast }\cap \mathcal{P}=\left\{ \frac{dQ}{d\mathbb{P}}\mid
Q\in (L_{\mathcal{F}})_{+}^{\ast }\text{ and }Q(1)=1\right\} $. Since $\pi
_{A}$ is $\sigma (L_{\mathcal{F}},L_{\mathcal{F}}^{\ast })$-lsc the
representation (\ref{piAsup}) follows immediately applying Proposition \ref%
{rapprR} to the map $\pi _{A}$ and observing that 
\begin{eqnarray*}
\pi _{A}(X) &=&\sup_{Q\in L_{\mathcal{F}}^{\ast }\cap \mathcal{P}}\inf_{\xi
\in L_{\mathcal{F}}}\left\{ \pi _{A}(\xi )\mid E_{Q}[\xi ]\geq
E_{Q}[X]\right\} \\
&\leq &\sup_{Q\in L_{\mathcal{F}}^{\ast }\cap \mathcal{P}}\inf_{\xi \in L_{%
\mathcal{F}}}\left\{ \pi _{A}(\xi )\mid E_{Q}[\xi |\mathcal{G}]\geq
_{Q}E_{Q}[X|\mathcal{G}]\right\} \leq \pi _{A}(X).
\end{eqnarray*}
\end{proof}

\section{Proof of the main results}

\label{proof}

Notations: In the following, we will only consider \textit{finite}
partitions $\Gamma =\left\{ A^{\Gamma }\right\} $ of $\mathcal{G} $\
measurable sets $A^{\Gamma }\in \Gamma $ and we set 
\begin{eqnarray*}
\pi ^{\Gamma }(X) &:&=\sum_{A^{\Gamma }\in \Gamma }\pi _{A^{\Gamma }}(X)%
\mathbf{1}_{A^{\Gamma }}, \\
K^{\Gamma }(X,Q) &:&=\inf_{\xi \in L_{\mathcal{F}}}\left\{ \pi ^{\Gamma
}(\xi )\mid E_{Q}[\xi |\mathcal{G}]\geq_Q E_{Q}[X|\mathcal{G}]\right\} \\
H^{\Gamma }(X) &:&=\sup_{Q\in L_{\mathcal{F}}^{\ast }\cap \mathcal{P}%
}K^{\Gamma }(X,Q)
\end{eqnarray*}

\subsection{Outline of the proof\label{Sec421}}

We anticipate an heuristic sketch of the proof of Theorem \ref{main},
pointing out the essential arguments involved in it and we defer to the
following section the details and the rigorous statements.

\bigskip

The proof relies on the equivalence of the following conditions:

\begin{enumerate}
\item $\pi (X)=H(X)$.

\item $\forall \,\varepsilon >0$, $\exists \,Q_{\varepsilon }\in L_{\mathcal{%
F}}^{\ast }\cap \mathcal{P}$ such that $\pi (X)-K(X,Q_{\varepsilon
})<\varepsilon $.

\item $\forall \,\varepsilon >0$, $\exists \,Q_{\varepsilon }\in L_{\mathcal{%
F}}^{\ast }\cap \mathcal{P}$ such that 
\begin{equation}
\{\xi \in L_{\mathcal{F}}\mid E_{Q_{\varepsilon }}[\xi |\mathcal{G}%
]\geq_{Q_{\varepsilon }} E_{Q_{\varepsilon }}[X|\mathcal{G}]\}\subseteq
\{\xi \in L_{\mathcal{F}}\mid \pi (\xi )>\pi (X)-\varepsilon \}.  \label{hb}
\end{equation}
\end{enumerate}

Indeed, $1.\Rightarrow 2.$ is a consequence of Proposition \ref{LQQ} (when
it holds true); $2.\Rightarrow 3.$ follows from the observation that $\pi
(X)<K(X,Q_{\varepsilon })+\varepsilon $ implies $\pi (X)<\pi (\xi
)+\varepsilon $ for every $\xi $ satisfying $E_{Q_{\varepsilon }}[\xi |%
\mathcal{G}]\geq_{Q_{\varepsilon }} E_{Q_{\varepsilon }}[X|\mathcal{G}]$; $%
3.\Rightarrow 1$. is implied by the inequalities: 
\begin{eqnarray*}
\pi (X)-\varepsilon &\leq &\inf \{\pi (\xi )\mid \pi (\xi )>\pi
(X)-\varepsilon \} \\
&\leq &\inf_{\xi \in L_{\mathcal{F}}}\{\pi (\xi )\mid E_{Q_{\varepsilon
}}[\xi |\mathcal{G}]\geq_{Q_{\varepsilon }} E_{Q_{\varepsilon }}[X|\mathcal{G%
}]\}\leq H(X)\leq \pi (X).
\end{eqnarray*}%
Unfortunately, we cannot prove Item 3. directly, relying on Hahn-Banach
Theorem, as it happened in the real case (see the proof of Theorem \ref%
{Volle1}, equation (\ref{finitecase}), in Appendix). Indeed, the complement
of the set in the RHS of (\ref{hb}) is not any more a convex set - unless $%
\pi $ is real valued - regardless of the continuity assumption made on $\pi $%
.

Also the method applied in the conditional convex case \cite{Sca} can not be
used here, since the map $X\rightarrow E_{\mathbb{P}}[\pi (X)]$ there
adopted preserves convexity but not quasiconvexity.

The idea is then to apply an approximation argument and the choice of
approximating $\pi (\cdot )$ by $\pi ^{\Gamma }(\cdot )$, is forced by the
need to preserve quasiconvexity.

\begin{enumerate}
\item[I] The first step is to prove (see Proposition \ref{propR}) that: $%
H^{\Gamma }(X)=\pi ^{\Gamma }(X).$ This is based on the representation of
the \textit{real valued} quasiconvex map $\pi _{A}$ in Proposition \ref%
{rapprPiA}. Therefore, the assumptions (LSC), (MON), (REG) and (QCO) on $\pi 
$ are here all needed.

\item[II] Then it is a simple matter to deduce $\pi (X)=\inf_{\Gamma }\pi
^{\Gamma }(X)=\inf_{\Gamma }H^{\Gamma }(X)$, where the $\inf $ is taken with
respect to all finite partitions.

\item[III] As anticipated in (\ref{last1}), the last step, i.e. proving that 
$\inf_{\Gamma }H^{\Gamma }(X)=H(X)$, is more delicate. It can be shown
easily that is possible to approximate $H(X)$ with $K(X,Q_{\varepsilon })$
on a set $A_{\varepsilon }$ of probability arbitrarily close to $1$.
However, we need the following \textit{uniform} approximation: For any$%
\,\varepsilon >0$ there exists $Q_{\varepsilon }\in L_{\mathcal{F}}^{\ast
}\cap \mathcal{P}$ such that for any finite partition $\Gamma $ we have $%
H^{\Gamma }(X)-K^{\Gamma }(X,Q_{\varepsilon })<\varepsilon $ on the same set 
$A_{\varepsilon }$. This key approximation result, based on Lemma \ref%
{maggiorazione}, shows that the element $Q_{\varepsilon }$ does not depend
on the partition and allows us (see equation (\ref{last})) to conclude the
proof .
\end{enumerate}

\subsection{Details\label{422}}

The following two lemmas are applications of measure theory

\begin{lemma}
\label{approx} For every $Y\in L_{\mathcal{G}}^{0}$ there exists a sequence $%
\Gamma (n)$ of finite partitions such that $\sum_{\Gamma (n)}\left(
\sup_{A^{\Gamma (n)}}Y\right) \mathbf{1}_{A^{\Gamma (n)}}$ converges in
probability, and $\mathbb{P} $-a.s., to $Y$.
\end{lemma}

\begin{proof}
Fix $\varepsilon ,\delta >0$ and consider the partitions $\Gamma
(n)=\{A_{0}^{n},A_{1}^{n},...A_{n2^{n+1}+1}^{n}\}$ where 
\begin{eqnarray*}
A_{0}^{n} &=&\{Y\in (-\infty ,-n]\} \\
A_{j}^{n} &=&\{Y\in (-n+\frac{j-1}{2^{n}},-n+\frac{j}{2^{n}}]\}\quad \forall
\,j=1,...,n2^{n+1} \\
A_{n2^{n+1}+1}^{n} &=&\{Y\in (n,+\infty )\}
\end{eqnarray*}%
Since $\mathbb{P}(A_{0}^{n}\cup A_{n2^{n+1}+1}^{n})\rightarrow 0$ as $%
n\rightarrow \infty ,$ we consider $N$ such that $\mathbb{P}(A_{0}^{N}\cup
A_{N2^{N}+1}^{N})\leq 1-\varepsilon $. Moreover we may find $M$ such that $%
\frac{1}{2^{M}}<\delta $, and hence for $\Gamma =\Gamma (M\vee N)$ we have: 
\begin{equation}
\mathbb{P}\left\{ \omega \in \Omega \mid \sum_{A^{\Gamma }\in \Gamma }\left(
\sup_{A^{\Gamma }}Y\right) \mathbf{1}_{A^{\Gamma }}(\omega )-Y(\omega
)<\delta \right\} >1-\varepsilon .  \label{MN}
\end{equation}
\end{proof}

\begin{lemma}
\label{LK}For each $X\in L_{\mathcal{F}}$ and $Q\in L_{\mathcal{F}}^{\ast
}\cap \mathcal{P}$ 
\begin{equation*}
\inf_{\Gamma }K^{\Gamma }(X,Q)=K(X,Q)
\end{equation*}%
where the \emph{infimum} is taken with respect to all finite partitions $%
\Gamma $.
\end{lemma}

\begin{proof}
\begin{eqnarray}
\inf_{\Gamma }K^{\Gamma }(X,Q) &=&\inf_{\Gamma }\inf_{\xi \in L_{\mathcal{F}%
}}\left\{ \pi ^{\Gamma }(\xi )\mid E_{Q}[\xi |\mathcal{G}]\geq_Q E_{Q}[X|%
\mathcal{G}]\right\}  \notag \\
&=&\inf_{\xi \in L_{\mathcal{F}}}\left\{ \inf_{\Gamma }\pi ^{\Gamma }(\xi
)\mid E_{Q}[\xi |\mathcal{G}]\geq_Q E_{Q}[X|\mathcal{G}]\right\}  \notag \\
&=&\inf_{\xi \in L_{\mathcal{F}}}\left\{ \pi (\xi )\mid E_{Q}[\xi |\mathcal{G%
}]\geq_Q E_{Q}[X|\mathcal{G}]\right\} =K(X,Q).  \label{Linf}
\end{eqnarray}%
where the first equality in (\ref{Linf}) follows from the convergence shown
in Lemma \ref{approx}.
\end{proof}

\bigskip

The following already mentioned key result is proved in the Appendix, for it
needs a pretty long argument.

\begin{lemma}
\label{maggiorazione}Let $X\in L_{\mathcal{F}}$ and let $P$ and $Q$ be
arbitrary elements of $L_{\mathcal{F}}^{\ast }\cap \mathcal{P}$. Suppose
that there exists $B\in \mathcal{G}$ satisfying: $K(X,P)\mathbf{1}%
_{B}>-\infty $, $\pi _{B}(X)<+\infty $ and 
\begin{equation*}
K(X,Q)\mathbf{1}_{B}\leq K(X,P)\mathbf{1}_{B}+\varepsilon \mathbf{1}_{B}%
\text{,}
\end{equation*}%
for some $\varepsilon \geq 0$. Then for every partition $\Gamma =\{B^{C},%
\widetilde{\Gamma }\}$, where $\widetilde{\Gamma }$ is a partition of $B$,
we have 
\begin{equation*}
K^{\Gamma }(X,Q)\mathbf{1}_{B}\leq K^{\Gamma }(X,P)\mathbf{1}%
_{B}+\varepsilon \mathbf{1}_{B}.
\end{equation*}
\end{lemma}

Since $\pi ^{\Gamma }$ assumes only a finite number of values, we may apply
Proposition \ref{rapprPiA} and deduce the dual representation of $\pi
^{\Gamma }$.

\begin{proposition}
\label{propR}Suppose that the assumptions of Theorem \ref{main} hold true
and $\Gamma $ is a finite partition. If for every $X\in L_{\mathcal{F}}$, $%
\pi ^{\Gamma }(X)<+\infty $ then: 
\begin{equation}
H^{\Gamma }(X)=\pi ^{\Gamma }(X)\geq \pi (X)  \label{Hn}
\end{equation}%
and therefore 
\begin{equation*}
\inf_{\Gamma }H^{\Gamma }(X)=\pi (X).
\end{equation*}
\end{proposition}

\begin{proof}
First notice that $K^{\Gamma }(X,Q)\leq H^{\Gamma }(X)\leq \pi ^{\Gamma
}(X)<+\infty $ for all $Q\in L_{\mathcal{F}}^{\ast }\cap \mathcal{P}$.
Consider the sigma algebra $\mathcal{G}^{\Gamma }:=\sigma (\Gamma )\subseteq 
\mathcal{G}$, generated by the finite partition $\Gamma $. Hence from
Proposition \ref{rapprPiA} we have for every $A^{\Gamma }\in \Gamma $ 
\begin{equation}
\pi _{A^{\Gamma }}(X)=\sup_{Q\in L_{\mathcal{F}}^{\ast }\cap \mathcal{P}%
}\inf_{\xi \in L_{\mathcal{F}}}\left\{ \pi _{A^{\Gamma }}(\xi )\mid
E_{Q}[\xi |\mathcal{G}]\geq_Q E_{Q}[X|\mathcal{G}]\right\} .  \label{130}
\end{equation}%
Moreover $H^{\Gamma }(X)$ is constant on $A^{\Gamma }$ since it is $\mathcal{%
G}^{\Gamma }$-measurable as well. Using the fact that $\pi ^{\Gamma }(\cdot
) $ is constant on each $A^{\Gamma }$, for every $A^{\Gamma }\in \Gamma $ we
then have: 
\begin{eqnarray}
H^{\Gamma }(X)\mathbf{1}_{A^{\Gamma }} &=&\sup_{Q\in L_{\mathcal{F}}^{\ast
}\cap \mathcal{P}}\inf_{\xi \in L_{\mathcal{F}}}\left\{ \pi ^{\Gamma }(\xi )%
\mathbf{1}_{A^{\Gamma }}\mid E_{Q}[\xi |\mathcal{G}]\geq_Q E_{Q}[X|\mathcal{G%
}]\right\}  \notag \\
&=&\sup_{Q\in L_{\mathcal{F}}^{\ast }\cap \mathcal{P}}\inf_{\xi \in L_{%
\mathcal{F}}}\left\{ \pi _{A^{\Gamma }}(\xi )\mathbf{1}_{A^{\Gamma }}\mid
E_{Q}[\xi |\mathcal{G}]\geq_Q E_{Q}[X|\mathcal{G}]\right\}  \notag \\
&=&\pi _{A^{\Gamma }}(X)\mathbf{1}_{A^{\Gamma }}=\pi ^{\Gamma }(X)\mathbf{1}%
_{A^{\Gamma }}  \label{120}
\end{eqnarray}%
where the first equality in (\ref{120}) follows from (\ref{130}). The
remaining statement is a consequence of (\ref{Hn}) and Lemma \ref{approx}
\end{proof}

\bigskip

\begin{proof}[Proof of Theorem \protect\ref{main}]
Obviously $\pi (X)\geq H(X),$ since $X$ satisfies the constraints in the
definition of $H(X)$. We may assume w.l.o.g. that $\pi (0)=0$ and so $\pi (X%
\mathbf{1}_{G})=\pi (X)\mathbf{1}_{G}$ for every $G\in \mathcal{G}$ (indeed,
otherwise we could consider $\rho (\cdot )=\pi (\cdot )-\pi (0)$ ). \newline
First we assume that $\pi $ is uniformly bounded, i.e. there exists $c>0$
such that for all $X\in L_{\mathcal{F}}$ $|\pi (X)|\leq c.$ Then $%
H(X)>-\infty $.

From Lemma \ref{LH}, eq. (\ref{lim2}), we know that there exists a sequence $%
Q_{k}\in L_{\mathcal{F}}^{\ast }\cap \mathcal{P}$ such that:%
\begin{equation*}
K(X,Q_{k})\uparrow H(X)\text{, as }k\uparrow \infty .
\end{equation*}%
Therefore, for any $\varepsilon >0$ we may find $Q_{\varepsilon }\in L_{%
\mathcal{F}}^{\ast }\cap \mathcal{P}$ and $A_{\varepsilon }\in \mathcal{G}$, 
$\mathbb{P}(A_{\varepsilon })>1-\varepsilon $ such that 
\begin{equation*}
H(X)\mathbf{1}_{A_{\varepsilon }}-K(X,Q_{\varepsilon })\mathbf{1}%
_{A_{\varepsilon }}\leq \varepsilon \mathbf{1}_{A_{\varepsilon }}.
\end{equation*}%
Since $H(X)\geq K(X,Q)$ $\forall Q\in L_{\mathcal{F}}^{\ast }\cap \mathcal{P}
$,%
\begin{equation*}
(K(X,Q_{\varepsilon })+\varepsilon )\mathbf{1}_{A_{\varepsilon }}\geq K(X,Q)%
\mathbf{1}_{A_{\varepsilon }}\text{ }\forall Q\in L_{\mathcal{F}}^{\ast
}\cap \mathcal{P}.
\end{equation*}%
This is the basic inequality that enable us to apply Lemma \ref%
{maggiorazione}, replacing there $P$ with $Q_{\varepsilon }$ and $B$ with $%
A_{\varepsilon }$. Only notice that $\sup_{\Omega }\pi (X)\leq c$ and $%
K(X,Q)>-\infty $ for every $Q\in L_{\mathcal{F}}^{\ast }\cap \mathcal{P}$.
This Lemma assures that for every partition $\Gamma $ of $\Omega $ 
\begin{equation}
(K^{\Gamma }(X,Q_{\varepsilon })+\varepsilon )\mathbf{1}_{A_{\varepsilon
}}\geq K^{\Gamma }(X,Q)\mathbf{1}_{A_{\varepsilon }}\text{ }\forall Q\in L_{%
\mathcal{F}}^{\ast }\cap \mathcal{P}\text{.}  \label{indotta}
\end{equation}%
From the definition of \emph{essential supremum} of a class of r.v. equation
(\ref{indotta}) implies that for every $\Gamma $ 
\begin{equation}
(K^{\Gamma }(X,Q_{\varepsilon })+\varepsilon )\mathbf{1}_{A_{\varepsilon
}}\geq \sup_{Q\in L_{\mathcal{F}}^{\ast }\cap \mathcal{P}}K^{\Gamma }(X,Q)%
\mathbf{1}_{A_{\varepsilon }}=H^{\Gamma }(X)\mathbf{1}_{A_{\varepsilon }}.
\label{last}
\end{equation}%
Since $\pi ^{\Gamma }\leq c$, applying Proposition \ref{propR}, equation (%
\ref{Hn}), we get 
\begin{equation*}
(K^{\Gamma }(X,Q_{\varepsilon })+\varepsilon )\mathbf{1}_{A_{\varepsilon
}}\geq \pi (X)\mathbf{1}_{A_{\varepsilon }}.
\end{equation*}%
Taking the \emph{infimum} over all possible partitions, as in Lemma \ref{LK}%
, we deduce: 
\begin{equation}
(K(X,Q_{\varepsilon })+\varepsilon )\mathbf{1}_{A_{\varepsilon }}\geq \pi (X)%
\mathbf{1}_{A_{\varepsilon }}.  \label{Ok}
\end{equation}%
Hence, for any $\varepsilon >0$%
\begin{equation*}
(K(X,Q_{\varepsilon })+\varepsilon )\mathbf{1}_{A_{\varepsilon }}\geq \pi (X)%
\mathbf{1}_{A_{\varepsilon }}\geq H(X)\mathbf{1}_{A_{\varepsilon }}\geq
K(X,Q_{\varepsilon })\mathbf{1}_{A_{\varepsilon }}
\end{equation*}%
which implies $\pi (X)=H(X)$, since $\mathbb{P}(A_{\varepsilon })\rightarrow
1$ as $\varepsilon\rightarrow 0$.

Now we consider the case when $\pi $ is not necessarily bounded. We define
the new map $\psi (\cdot ):=\arctan (\pi (\cdot ))$ and notice that $\psi
(\xi )$ is a $\mathcal{G}$-measurable r.v. satisfying $|\psi (X)|\leq \frac{%
\Pi }{2}$ for every $X\in L_{\mathcal{F}}$. Moreover $\psi $ is (MON),
(QCO), (LSC) and $\psi (X\mathbf{1}_{G})=\psi (X)\mathbf{1}_{G}$ for every $%
G\in \mathcal{G}$. Since $\psi $ is bounded, by the above argument we may
conclude that 
\begin{equation*}
\psi (X)=H_{\psi }(X):=\sup_{Q\in L_{\mathcal{F}}^{\ast }\cap \mathcal{P}%
}K_{\psi }(X,Q)
\end{equation*}%
where 
\begin{equation*}
K_{\psi }(X,Q):=\inf_{\xi \in L_{\mathcal{F}}}\left\{ \psi (\xi )\mid
E_{Q}[\xi |\mathcal{G}]\geq _{Q}E_{Q}[X|\mathcal{G}]\right\} .
\end{equation*}%
Applying again Lemma \ref{LH}, equation (\ref{lim2}), there exists $Q^{k}\in
L_{\mathcal{F}}^{\ast }$ such that%
\begin{equation*}
H_{\psi }(X)=\lim_{k}K_{\psi }(X,Q^{k}).
\end{equation*}%
We will show below that 
\begin{equation}
K_{\psi }(X,Q^{k})=\arctan K(X,Q^{k}).  \label{arc}
\end{equation}%
Admitting this, we have for $\mathbb{P}$-almost every $\omega \in \Omega $%
\begin{eqnarray*}
\arctan (\pi (X)(\omega )) &=&\psi (X)(\omega )=H_{\psi }(X)(\omega
)=\lim_{k}K_{\psi }(X,Q^{k})(\omega ) \\
&=&\lim_{k}\arctan K(X,Q^{k})(\omega ))=\arctan (\lim_{k}K(X,Q^{k})(\omega
)),
\end{eqnarray*}%
where we used the continuity of the function $\arctan $. This implies $\pi
(X)=\lim_{k}K(X,Q^{k})$ and we conclude: 
\begin{equation*}
\pi (X)=\lim_{k}K(X,Q^{k})\leq H(X)\leq \pi (X).
\end{equation*}%
It only remains to show (\ref{arc}). We prove that for every fixed $Q\in L_{%
\mathcal{F}}^{\ast }\cap \mathcal{P}$ 
\begin{equation*}
K_{\psi }(X,Q)=\arctan \left( K(X,Q)\right) .
\end{equation*}%
Since $\pi $ and $\psi $ are regular, from Lemma \ref{upwardn} iv), there
exist $\xi _{h}^{Q}\in L_{\mathcal{F}}$ and $\eta _{h}^{Q}\in L_{\mathcal{F}%
} $ such that 
\begin{equation}
E_{Q}[\xi _{h}^{Q}|\mathcal{G}]\geq _{Q}E_{Q}[X|\mathcal{G}]\text{, }%
E_{Q}[\eta _{h}^{Q}|\mathcal{G}]\geq _{Q}E_{Q}[X|\mathcal{G}]\text{, }%
\forall h\geq 1,  \label{fi}
\end{equation}%
$\psi (\xi _{h}^{Q})\downarrow K_{\psi }(X,Q)$ and $\pi (\eta
_{h}^{Q})\downarrow K(X,Q)$, as $h\uparrow \infty $. From (\ref{fi}) and the
definitions of $K(X,Q)$, $K_{\psi }(X,Q)$ and by the continuity and
monotonicity of $\arctan $ we get: 
\begin{eqnarray*}
K_{\psi }(X,Q) &\leq &\lim_{h}\psi (\eta _{h}^{Q})=\lim_{h}\arctan \pi (\eta
_{h}^{Q})=\arctan \lim_{h}\pi (\eta _{h}^{Q}) \\
&=&\arctan K(X,Q)\leq \arctan \lim_{h}\pi (\xi _{h}^{Q})=\lim_{h}\psi (\xi
_{h}^{Q})=K_{\psi }(X,Q).
\end{eqnarray*}
\end{proof}

\begin{remark}
\label{conservaAtteso}Consider $Q\in \mathcal{P}$ such that $Q\sim \mathbb{P}
$ on $\mathcal{G}$ and define the new probability 
\begin{equation*}
\widetilde{Q}(F):=E_{Q}\left[ \frac{d\mathbb{P}}{dQ}^{\mathcal{G}}\mathbf{1}%
_{F}\right] \quad \text{where}\quad \frac{d\mathbb{P}}{dQ}^{\mathcal{G}%
}=:E_{Q}\left[ \frac{d\mathbb{P}}{dQ}\big|\,\mathcal{G}\right] ,\text{ }F\in 
\mathcal{F}.
\end{equation*}%
Then $\widetilde{Q}(G)=\mathbb{P}(G)$ for all $G\in \mathcal{G}$, and so $%
\widetilde{Q}\in \mathcal{P}_{\mathcal{G}}$. Moreover, it is easy to check
that for all $X\in L_{\mathcal{F}}$ and $Q\in L_{\mathcal{F}}^{\ast }\cap 
\mathcal{P}$ such that $Q\sim \mathbb{P}$ on $\mathcal{G}$ we have: 
\begin{equation*}
E_{\widetilde{Q}}[X|\mathcal{G}]=E_{Q}[X|\mathcal{G}]
\end{equation*}%
which implies $K(X,\widetilde{Q})=K(X,Q).$
\end{remark}

\begin{proof}[Proof of Corollary \protect\ref{Cor1}]
Consider the probability $Q_{\varepsilon }\in L_{\mathcal{F}}^{\ast }\cap 
\mathcal{P}$ built up in Theorem \ref{main}, equation (\ref{Ok}). We claim
that $Q_{\varepsilon }$ is equivalent to $\mathbb{P}$ on $A_{\varepsilon }$.
By contradiction there exists $B\in \mathcal{G}$, $B\subseteq A_{\varepsilon
}$, such that $\mathbb{P}(B)>0$ but $Q_{\varepsilon }(B)=0$. Consider $\eta
\in L_{\mathcal{F}}$, $\delta >0$ such that $\mathbb{P}(\pi (\eta )+\delta
<\pi (X))=1$ and define $\xi =X\mathbf{1}_{B^{C}}+\eta \mathbf{1}_{B}$ so
that $E_{Q_{\varepsilon }}[\xi |\mathcal{G}]\geq_{Q_{\varepsilon }}
E_{Q_{\varepsilon }}[X|\mathcal{G}]$. By regularity $\pi (\xi )=\pi (X)%
\mathbf{1}_{B^{C}}+\pi (\eta )\mathbf{1}_{B}$ which implies for $\mathbb{P}$%
-a.e. $\omega \in B$ 
\begin{equation*}
\pi (\xi )(\omega )+\delta =\pi (\eta )(\omega )+\delta <\pi (X)(\omega
)\leq K(X,Q_{\varepsilon })(\omega )+\varepsilon \leq \pi (\xi )(\omega
)+\varepsilon \text{ }
\end{equation*}%
which is impossible for $\varepsilon \leq \delta .$ So $Q_{\varepsilon }\sim 
\mathbb{P}$ on $A_{\varepsilon }$ for all small $\varepsilon \leq \delta $. 
\newline
Consider $\widehat{Q}_{\varepsilon }$ such that $\frac{d\widehat{Q}%
_{\varepsilon }}{d\mathbb{P}}=\frac{dQ_{\varepsilon }}{d\mathbb{P}}\mathbf{1}%
_{A_{\varepsilon }}+\frac{d\mathbb{P}}{d\mathbb{P}}\mathbf{1}%
_{(A_{\varepsilon })^{C}}$. Up to a normalization factor $\widehat{Q}%
_{\varepsilon }\in L_{\mathcal{F}}^{\ast }\cap \mathcal{P}$ and is
equivalent to $\mathbb{P}$. Moreover from Lemma \ref{upwardn} (vi), $K(X,%
\widehat{Q}_{\varepsilon })\mathbf{1}_{A_{\varepsilon }}=K(X,Q_{\varepsilon
})\mathbf{1}_{A_{\varepsilon }}$ and from Remark \ref{conservaAtteso} we may
define $\widetilde{Q}_{\varepsilon }\in \mathcal{P}_{\mathcal{G}}$ such that 
$K(X,\widetilde{Q}_{\varepsilon })\mathbf{1}_{A_{\varepsilon }}=K(X,\widehat{%
Q}_{\varepsilon })\mathbf{1}_{A_{\varepsilon }}=K(X,Q_{\varepsilon })\mathbf{%
1}_{A_{\varepsilon }}$. From (\ref{Ok}) we finally deduce: $K(X,\widetilde{Q}%
_{\varepsilon })\mathbf{1}_{A_{\varepsilon }}+\varepsilon \mathbf{1}%
_{A_{\varepsilon }}\geq \pi (X)\mathbf{1}_{A_{\varepsilon }}$, and the
thesis then follows from $\widetilde{Q}_{\varepsilon }\in \mathcal{P}_{%
\mathcal{G}}$.
\end{proof}

\section{Appendix}

\subsection{Proof of the key approximation Lemma \protect\ref{maggiorazione} 
\label{a51}}

We will adopt the following notations: If $\Gamma _{1}$ and $\Gamma _{2}$
are two finite partitions of $\mathcal{G}$-measurable sets then $\Gamma
_{1}\cap \Gamma _{2}:=\left\{ A_{1}\cap A_{2}\mid A_{i}\in \Gamma
_{i},\;i=1,2\right\} $ is a finite partition finer than each $\Gamma _{1}$
and $\Gamma _{2}$.

Lemma \ref{down1} is the natural generalization of Lemma \ref{down} to the
approximated problem.

\begin{lemma}
\label{down1}For every partition $\Gamma $, $X\in L_{\mathcal{F}}$ and $Q\in
L_{\mathcal{F}}^{\ast }\cap \mathcal{P}$, the set 
\begin{equation*}
\mathcal{A}_{Q}^{\Gamma }(X)\circeq \{\pi ^{\Gamma }(\xi )\,|\,\xi \in L_{%
\mathcal{F}}\;\text{and }E_{Q}[\xi |\mathcal{G}]\geq_Q E_{Q}[X|\mathcal{G}]\}
\end{equation*}%
is downward directed. This implies that there exists exists a sequence $%
\left\{ \eta _{m}^{Q}\right\} _{m=1}^{\infty }\in L_{\mathcal{F}}$ such that 
\begin{equation*}
E_{Q}[\eta _{m}^{Q}|\mathcal{G}]\geq_Q E_{Q}[X|\mathcal{G}]\text{ }\forall
m\geq 1\text{ , }\quad \pi ^{\Gamma }(\eta _{m}^{Q})\downarrow K^{\Gamma
}(X,Q)\text{ as }m\uparrow \infty .
\end{equation*}
\end{lemma}

\begin{proof}
To show that the set $\mathcal{A}_{Q}^{\Gamma }(X)$ is downward directed we
use the notations and the results in the proof of Lemma \ref{down} and check
that 
\begin{equation*}
\pi ^{\Gamma }(\xi ^{\ast })=\pi ^{\Gamma }(\xi _{1}\mathbf{1}_{G}+\xi _{2}%
\mathbf{1}_{G^{C}})\leq \min \left\{ \pi ^{\Gamma }(\xi _{1}),\pi ^{\Gamma
}(\xi _{2})\right\} .
\end{equation*}
\end{proof}

\bigskip

Now we show that for any given sequence of partition there exists one
sequence that works for all.

\begin{lemma}
\label{LQ}For any fixed, at most countable, family of partitions $\{\Gamma
(h)\}_{h\geq 1}$ and $Q\in L_{\mathcal{F}}^{\ast }\cap \mathcal{P},$ there
exists a sequence $\left\{ \xi _{m}^{Q}\right\} _{m=1}^{\infty }\in L_{%
\mathcal{F}}$ such that 
\begin{eqnarray*}
E_{Q}[\xi _{m}^{Q}|\mathcal{G}] &\geq_Q &E_{Q}[X|\mathcal{G}]\quad \text{
for all }m\geq 1 \\
\pi (\xi _{m}^{Q}) &\downarrow &K(X,Q)\quad \text{ as }m\uparrow \infty \\
\text{and for all }h\quad \pi ^{\Gamma (h)}(\xi _{m}^{Q}) &\downarrow
&K^{\Gamma (h)}(X,Q)\quad \text{ as }m\uparrow \infty .
\end{eqnarray*}
\end{lemma}

\begin{proof}
Apply Lemma \ref{down} and Lemma \ref{down1} and find $\{\varphi
_{m}^{0}\}_{m},\{\varphi _{m}^{1}\}_{m},...,\{\varphi _{m}^{h}\}_{m},...$
such that for every $i$ and $m$ we have $E_{Q}[\varphi _{m}^{i}\mid \mathcal{%
G}]\geq_Q E_{Q}[X|\mathcal{G}]$ and 
\begin{eqnarray*}
\pi (\varphi _{m}^{0}) &\downarrow &K(X,Q)\quad \text{ as }m\uparrow \infty
\\
\text{and for all }h\quad \pi ^{\Gamma (h)}(\varphi _{m}^{h}) &\downarrow
&K^{\Gamma (h)}(X,Q)\quad \text{ as }m\uparrow \infty .
\end{eqnarray*}%
For each $m\geq 1$ consider $\bigwedge_{i=0}^{m}\pi (\varphi _{m}^{i})$:
then there will exists a (non unique) finite partition of $\Omega $, $%
\{F_{m}^{i}\}_{i=1}^{m}$ such that 
\begin{equation*}
\bigwedge_{i=0}^{m}\pi (\varphi _{m}^{i})=\sum_{i=0}^{m}\pi (\varphi
_{m}^{i})\mathbf{1}_{F_{m}^{i}}.
\end{equation*}%
Denote $\xi _{m}^{Q}=:\sum_{i=0}^{m}\varphi _{m}^{i}\mathbf{1}_{F_{m}^{i}}$
and notice that $\sum_{i=0}^{m}\pi (\varphi _{m}^{i})\mathbf{1}_{F_{m}^{i}}%
\overset{(REG)}{=}\pi \left( \xi _{m}^{Q}\right) $ and $E_{Q}[\xi _{m}^{Q}|%
\mathcal{G}]\geq_Q E_{Q}[X|\mathcal{G}]$ for every $m$. Moreover $\pi (\xi
_{m}^{Q})$ is decreasing and $\pi (\xi _{m}^{Q})\leq \pi (\varphi _{m}^{0})$
implies $\pi (\xi _{m}^{Q})\downarrow K(X,Q)$. \newline
For every fixed $h$ we have $\pi (\xi _{m}^{Q})\leq \pi (\varphi _{m}^{h})$
for all $h\leq m$ and hence: 
\begin{equation*}
\pi ^{\Gamma (h)}(\xi _{m}^{Q})\leq \pi ^{\Gamma (h)}(\varphi _{m}^{h})\text{
implies }\pi ^{\Gamma (h)}(\xi _{m}^{Q})\downarrow K^{\Gamma (h)}(X,Q)\text{
as }m\uparrow \infty .
\end{equation*}
\end{proof}

\bigskip

Finally, we state the basic step used in the proof of Lemma \ref%
{maggiorazione}.

\begin{lemma}
\label{Lnew}Let $X\in L_{\mathcal{F}}$ and let $P$ and $Q$ be arbitrary
elements of $L_{\mathcal{F}}^{\ast }\cap \mathcal{P}$. Suppose that there
exists $B\in \mathcal{G}$ satisfying: $K(X,P)\mathbf{1}_{B}>-\infty $, $\pi
_{B}(X)<+\infty $ and 
\begin{equation*}
K(X,Q)\mathbf{1}_{B}\leq K(X,P)\mathbf{1}_{B}+\varepsilon \mathbf{1}_{B}%
\text{,}
\end{equation*}%
for some $\varepsilon \geq 0$. Then for any $\delta >0$ and any partition $%
\Gamma _{0}$ there exists $\Gamma \supseteq \Gamma _{0}$ for which 
\begin{equation*}
K^{\Gamma }(X,Q)\mathbf{1}_{B}\leq K^{\Gamma }(X,P)\mathbf{1}%
_{B}+\varepsilon \mathbf{1}_{B}+\delta \mathbf{1}_{B}
\end{equation*}
\end{lemma}

\begin{proof}
By our assumptions we have: $-\infty <K(X,P)\mathbf{1}_{B}\leq \pi
_{B}(X)<+\infty $ and $K(X,Q)\mathbf{1}_{B}\leq \pi _{B}(X)<+\infty $. Fix $%
\delta >0$ and the partition $\Gamma _{0}$. Suppose by contradiction that
for any $\Gamma \supseteq \Gamma _{0}$ we have $\mathbb{P}(C)>0$ where 
\begin{equation}
C=\{\omega \in B\mid K^{\Gamma }(X,Q)(\omega )>K^{\Gamma }(X,P)(\omega
)+\varepsilon +\delta \}.  \label{kkk}
\end{equation}%
Notice that $C$ is the union of a finite number of elements in the partition 
$\Gamma $.

Consider that Lemma \ref{upwardn} guarantees the existence of $\left\{ \xi
_{h}^{Q}\right\} _{h=1}^{\infty }\in L_{\mathcal{F}}$ satisfying:%
\begin{eqnarray}
\pi (\xi _{h}^{Q}) &\downarrow &K(X,Q),\text{ as }h\uparrow \infty ,\text{ } 
\text{ ,}\quad E_{Q}[\xi _{h}^{Q}|\mathcal{G}] \geq_Q E_{Q}[X|\mathcal{G}]%
\text{ }\forall h\geq 1.  \label{Q1}
\end{eqnarray}%
Moreover, for each partition $\Gamma $ and $h\geq 1$ define: 
\begin{equation*}
D_{h}^{\Gamma }:=\left\{ \omega \in \Omega \mid \pi ^{\Gamma }(\xi
_{h}^{Q})(\omega )-\pi (\xi _{h}^{Q})(\omega )<\frac{\delta }{4}\right\} \in 
\mathcal{G},
\end{equation*}
and observe that $\pi ^{\Gamma }(\xi _{h}^{Q})$ decreases if we pass to
finer partitions. From Lemma \ref{approx} equation (\ref{MN}), we deduce
that for each $h\geq 1$ there exists a partition $\widetilde{\Gamma }(h)$
such that $\mathbb{P}\left( D_{h}^{\widetilde{\Gamma }(h)}\right) \geq 1-%
\frac{1}{2^{h}}$. For every $h\geq 1$ define the new partition $\Gamma
(h)=\left( \bigcap\limits_{j=1}^{h}\widetilde{\Gamma }(h)\right) \cap \Gamma
_{0}$ so that for all $h\geq 1$ we have: $\Gamma (h+1)\supseteq \Gamma
(h)\supseteq \Gamma _{0}$, $\mathbb{P}\left( D_{h}^{\Gamma (h)}\right) \geq
1-\frac{1}{2^{h}}$ and 
\begin{equation}
\left( \pi (\xi _{h}^{Q})+\frac{\delta }{4}\right) \mathbf{1}_{D_{h}^{\Gamma
(h)}}\geq \left( \pi ^{\Gamma (h)}(\xi _{h}^{Q})\right) \mathbf{1}%
_{D_{h}^{\Gamma (h)}}\text{, }\forall h\geq 1.  \label{BBB}
\end{equation}%
Lemma \ref{LQ} guarantees that for the fixed sequence of partitions $%
\{\Gamma (h)\}_{h\geq 1}$, there exists a sequence $\left\{ \xi
_{m}^{P}\right\} _{m=1}^{\infty }\in L_{\mathcal{F}}$, which does not depend
on $h$, satisfying 
\begin{eqnarray}
E_{P}[\xi _{m}^{P}|\mathcal{G}] &\geq_P &E_{P}[X|\mathcal{G}]\text{ }\forall
m\geq 1,  \label{32} \\
\pi ^{\Gamma (h)}(\xi _{m}^{P}) &\downarrow &K^{\Gamma (h)}(X,P),\text{ as }%
m\uparrow \infty ,\quad \forall \,h\geq 1.  \label{nnn}
\end{eqnarray}%
For each $m\geq 1$ and $\Gamma (h)$ define: 
\begin{equation*}
C_{m}^{\Gamma (h)}:=\left\{ \omega \in C\mid \pi ^{\Gamma (h)}(\xi
_{m}^{P})(\omega )-K^{\Gamma (h)}(X,P)(\omega )\leq \frac{\delta }{4}%
\right\} \in \mathcal{G}.
\end{equation*}%
Since the expressions in the definition of $C_{m}^{\Gamma (h)}$ assume only
a finite number of values, from (\ref{nnn}) and from our assumptions, which
imply that $K^{\Gamma (h)}(X,P)\geq K(X,P)>-\infty $ on $B$, we deduce that
for each $\Gamma (h)$ there exists an index $m(\Gamma (h))$ such that: $%
\mathbb{P}\left( C\setminus C_{m(\Gamma (h))}^{\Gamma (h)}\right) =0$ and 
\begin{equation}
K^{\Gamma (h)}(X,P)\mathbf{1}_{C_{m(\Gamma (h))}^{\Gamma (h)}}\geq \left(
\pi ^{\Gamma (h)}(\xi _{m(\Gamma (h))}^{P})-\frac{\delta }{4}\right) \mathbf{%
1}_{C_{m(\Gamma (h))}^{\Gamma (h)}}\text{, }\forall \,h\geq 1.  \label{CCC}
\end{equation}%
\newline
Set $E_{h}=D_{h}^{\Gamma (h)}\cap C_{m(\Gamma (h))}^{\Gamma (h)}$ $\in 
\mathcal{G}$ and observe that 
\begin{equation}
\mathbf{1}_{E_{h}}\rightarrow \mathbf{1}_{C}\quad \mathbb{P}-\text{a.s.}
\label{insieme}
\end{equation}%
From (\ref{BBB}) and (\ref{CCC}) we then deduce:%
\begin{eqnarray}
\left( \pi (\xi _{h}^{Q})+\frac{\delta }{4}\right) \mathbf{1}_{E_{h}} &\geq
&\left( \pi ^{\Gamma (h)}(\xi _{h}^{Q})\right) \mathbf{1}_{E_{h}},\text{ }%
\forall h\geq 1\text{,}  \label{Eh} \\
K^{\Gamma (h)}(X,P)\mathbf{1}_{E_{h}} &\geq &\left( \pi ^{\Gamma (h)}(\xi
_{m(\Gamma (h))}^{P})-\frac{\delta }{4}\right) \mathbf{1}_{E_{h}},\text{ }%
\forall h\geq 1.  \label{nh}
\end{eqnarray}%
We then have for any $h\geq 1$%
\begin{eqnarray}
\pi (\xi _{h}^{Q})\mathbf{1}_{E_{h}}+\frac{\delta }{4}\mathbf{1}_{E_{h}}
&\geq &\left( \pi ^{\Gamma (h)}(\xi _{h}^{Q})\right) \mathbf{1}_{E_{h}}
\label{1a} \\
&\geq &K^{\Gamma (h)}(X,Q)\mathbf{1}_{E_{h}}  \label{2a} \\
&\geq &\left( K^{\Gamma (h)}(X,P)+\varepsilon +\delta \right) \mathbf{1}%
_{E_{h}}  \label{3a} \\
&\geq &\left( \pi ^{\Gamma (h)}(\xi _{m(\Gamma (h))}^{P})-\frac{\delta }{4}%
+\varepsilon +\delta \right) \mathbf{1}_{E_{h}}  \label{4a} \\
&\geq &\left( \pi (\xi _{m(\Gamma (h))}^{P})+\varepsilon +\frac{3}{4}\delta
\right) \mathbf{1}_{E_{h}}.  \label{5a}
\end{eqnarray}%
(in the above chain of inequalities, (\ref{1a}) follows from (\ref{Eh}); (%
\ref{2a}) follows from (\ref{Q1}) and the definition of $K^{\Gamma (h)}(X,Q)$%
; (\ref{3a}) follows from (\ref{kkk}); (\ref{4a}) follows from (\ref{nh}); (%
\ref{5a}) follows from the definition of the maps $\pi _{A^{\Gamma (h)}}$).

\noindent Recalling (\ref{32}) we then get, for each $h\geq 1$,%
\begin{equation}
\pi (\xi _{h}^{Q})\mathbf{1}_{E_{h}}\geq \left( \pi (\xi _{m(\Gamma
(h))}^{P})+\varepsilon +\frac{\delta }{2}\right) \mathbf{1}_{E_{h}}\geq
\left( K(X,P)+\varepsilon +\frac{\delta }{2}\right) \mathbf{1}%
_{E_{h}}>-\infty .  \label{pdelta}
\end{equation}%
From equation (\ref{Q1}) and (\ref{insieme}) we have $\pi (\xi _{h}^{Q})%
\mathbf{1}_{E_{h}}\rightarrow K(X,Q)\mathbf{1}_{C}$ $\mathbb{P}$-a.s. as $%
h\uparrow \infty $ and so from (\ref{pdelta}) 
\begin{eqnarray*}
\mathbf{1}_{C}K(X,Q) &=&\lim_{h}\pi (\xi _{h}^{Q})\mathbf{1}_{E_{h}}\geq
\lim_{h}\mathbf{1}_{E_{h}}\left( K(X,P)+\varepsilon +\frac{\delta }{2}\right)
\\
&=&\mathbf{1}_{C}\left( K(X,P)+\varepsilon +\frac{\delta }{2}\right)
\end{eqnarray*}%
which contradicts the assumption of the Lemma, since $C\subseteq B$ and $%
\mathbb{P}(C)>0$.
\end{proof}

\bigskip

\begin{proof}[Proof of Lemma \protect\ref{maggiorazione}]
First notice that the assumptions of this Lemma are those of Lemma \ref{Lnew}%
. Assume by contradiction that there exists $\Gamma _{0}=\{B^{C},\widetilde{%
\Gamma }_{0}\},$ where $\widetilde{\Gamma }_{0}$ is a partition of $B$, such
that 
\begin{equation}
\mathbb{P(}\omega \in B\mid K^{\Gamma _{0}}(X,Q)(\omega )>K^{\Gamma
_{0}}(X,P)(\omega )+\varepsilon )>0.  \label{989}
\end{equation}%
By our assumptions we have $K^{\Gamma _{0}}(X,P)\mathbf{1}_{B}\geq K(X,P)%
\mathbf{1}_{B}>-\infty $ and $K^{\Gamma _{0}}(X,Q)\mathbf{1}_{B}\leq \pi
_{B}(X)\mathbf{1}_{B}<+\infty $. Since $K^{\Gamma _{0}}$ is constant on
every element $A^{\Gamma _{0}}\in \Gamma _{0},$ we denote with $K^{A^{\Gamma
_{0}}}(X,Q)$ the value that the random variable $K^{\Gamma _{0}}(X,Q)$
assumes on $A^{\Gamma _{0}}$. From (\ref{989}) we deduce that there exists $%
\widehat{A}^{\Gamma _{0}}\subseteq B$ , $\widehat{A}^{\Gamma _{0}}\in \Gamma
_{0}$, such that 
\begin{equation*}
+\infty >K^{\widehat{A}^{\Gamma _{0}}}(X,Q)>K^{\widehat{A}^{\Gamma
_{0}}}(X,P)+\varepsilon >-\infty .
\end{equation*}%
Let then $d>0$ be defined by 
\begin{equation}
d=:K^{\widehat{A}^{\Gamma _{0}}}(X,Q)-K^{\widehat{A}^{\Gamma
_{0}}}(X,P)-\varepsilon .  \label{bar}
\end{equation}%
Apply Lemma \ref{Lnew} with $\delta =\frac{d}{3}$: then there exists $\Gamma
\supseteq \Gamma _{0}$ (w.l.o.g. $\Gamma =\{B^{C},\widetilde{\Gamma }\}$
where $\widetilde{\Gamma }\supseteq \widetilde{\Gamma }_{0}$) such that 
\begin{equation}
K^{\Gamma }(X,Q)\mathbf{1}_{B}\leq \left( K^{\Gamma }(X,P)+\varepsilon
+\delta \right) \mathbf{1}_{B}.  \label{new}
\end{equation}%
Considering only the two partitions $\Gamma $ and $\Gamma _{0}$, we may
apply Lemma \ref{LQ} and conclude that there exist two sequences $\{\xi
_{h}^{P}\}_{h=1}^{\infty }\in L_{\mathcal{F}}$ and $\{\xi
_{h}^{Q}\}_{h=1}^{\infty }\in L_{\mathcal{F}}$ satisfying as $h\uparrow
\infty $: 
\begin{eqnarray}
E_{P}[\xi _{h}^{P}|\mathcal{G}]\geq_P E_{P}[X|\mathcal{G}], &\text{ } \pi
^{\Gamma _{0}}(\xi _{h}^{P})\downarrow K^{\Gamma _{0}}(X,P),&\text{ } \pi
^{\Gamma }(\xi _{h}^{P})\downarrow K^{\Gamma }(X,P)  \label{Q6} \\
E_{Q}[\xi _{h}^{Q}|\mathcal{G}]\geq_Q E_{Q}[X|\mathcal{G}], &\text{ } \pi
^{\Gamma _{0}}(\xi _{h}^{Q})\downarrow K^{\Gamma _{0}}(X,Q),&\text{ } \pi
^{\Gamma }(\xi _{h}^{Q})\downarrow K^{\Gamma }(X,Q)  \label{Q5}
\end{eqnarray}%
Since $K^{\Gamma _{0}}(X,P)$ is constant and finite on $\widehat{A}^{\Gamma
_{0}}$, from (\ref{Q6}) we may find $h_{1}\geq 1$ such that 
\begin{equation}
\pi _{\widehat{A}^{\Gamma _{0}}}(\xi _{h}^{P})-K^{\widehat{A}^{\Gamma
_{0}}}(X,P)<\frac{d}{2}\,,\forall \,h\geq h_{1}.  \label{vicino}
\end{equation}%
From equation (\ref{bar}) and (\ref{vicino}) we deduce that 
\begin{equation*}
\pi _{\widehat{A}^{\Gamma _{0}}}(\xi _{h}^{P})<K^{\widehat{A}^{\Gamma
_{0}}}(X,P)+\frac{d}{2}=K^{\widehat{A}^{\Gamma _{0}}}(X,Q)-\varepsilon -d+%
\frac{d}{2},\text{ }\forall \,h\geq h_{1},
\end{equation*}%
and therefore, knowing from (\ref{Q5}) that $K^{\widehat{A}^{\Gamma
_{0}}}(X,Q)\leq \pi _{\widehat{A}^{\Gamma _{0}}}(\xi _{h}^{Q}),$ 
\begin{equation}
\pi _{\widehat{A}^{\Gamma _{0}}}(\xi _{h}^{P})+\frac{d}{2}<\pi _{\widehat{A}%
^{\Gamma _{0}}}(\xi _{h}^{Q})-\varepsilon \quad \forall \,h\geq h_{1}.
\label{first}
\end{equation}%
We now take into account all the sets $A^{\Gamma }\subseteq \widehat{A}%
^{\Gamma _{0}}\subseteq B$. For the convergence of $\pi _{A^{\Gamma }}(\xi
_{h}^{Q})$ we distinguish two cases. On those sets $A^{\Gamma }$ for which $%
K^{A^{\Gamma }}(X,Q)>-\infty $ we may find, from (\ref{Q5}), $\overline{h}%
\geq 1$ such that 
\begin{equation*}
\pi _{A^{\Gamma }}(\xi _{h}^{Q})-K^{A^{\Gamma }}(X,Q)<\frac{\delta }{2}\text{
}\forall \,h\geq \overline{h}.
\end{equation*}%
Then using (\ref{new}) and (\ref{Q6}) we have 
\begin{equation*}
\pi _{A^{\Gamma }}(\xi _{h}^{Q})<K^{A^{\Gamma }}(X,Q)+\frac{\delta }{2}\leq
K^{A^{\Gamma }}(X,P)+\varepsilon +\delta +\frac{\delta }{2}\leq \pi
_{A^{\Gamma }}(\xi _{h}^{P})+\varepsilon +\delta +\frac{\delta }{2}
\end{equation*}%
so that 
\begin{equation*}
\pi _{A^{\Gamma }}(\xi _{h}^{Q})<\pi _{A^{\Gamma }}(\xi
_{h}^{P})+\varepsilon +\frac{3\delta }{2}\quad \forall \,h\geq \overline{h}.
\end{equation*}%
On the other hand, on those sets $A^{\Gamma }$ for which $K^{A^{\Gamma
}}(X,Q)=-\infty $ the convergence (\ref{Q5}) guarantees the existence of $%
\widehat{h}\geq 1$ for which we obtain again: 
\begin{equation}
\pi _{A^{\Gamma }}(\xi _{h}^{Q})<\pi _{A^{\Gamma }}(\xi
_{h}^{P})+\varepsilon +\frac{3\delta }{2}\quad \forall \,h\geq \widehat{h}
\label{second}
\end{equation}%
(notice that $K^{\Gamma }(X,P)\geq K(X,P)\mathbf{1}_{B}>-\infty $ and (\ref%
{Q6}) imply that $\pi _{A^{\Gamma }}(\xi _{h}^{P})$ converges to a finite
value, for $A^{\Gamma }\subseteq B$).

Since the partition $\Gamma $ is finite there exists $h_{2}\geq 1$ such that
equation (\ref{second}) stands for every $A^{\Gamma }\subseteq \widehat{A}%
^{\Gamma _{0}}$ and for every $h\geq h_{2}$ and for our choice of $\delta =%
\frac{d}{3}$ (\ref{second}) becomes 
\begin{equation}
\pi _{A^{\Gamma }}(\xi _{h}^{Q})<\pi _{A^{\Gamma }}(\xi
_{h}^{P})+\varepsilon +\frac{d}{2}\quad \forall \,h\geq h_{2}\quad \forall
\,A^{\Gamma }\subseteq \widehat{A}^{\Gamma _{0}}.  \label{second1}
\end{equation}%
\newline
Fix $h^{\ast }>\max \{h_{1},h_{2}\}$ and consider the value $\pi _{\widehat{A%
}^{\Gamma _{0}}}(\xi _{h^{\ast }}^{Q})$. Then among all $A^{\Gamma
}\subseteq \widehat{A}^{\Gamma _{0}}$ we may find $B^{\Gamma }\subseteq 
\widehat{A}^{\Gamma _{0}}$ such that $\pi _{B^{\Gamma }}(\xi _{h^{\ast
}}^{Q})=\pi _{\widehat{A}^{\Gamma _{0}}}(\xi _{h^{\ast }}^{Q}).$ Thus: 
\begin{equation*}
\pi _{\widehat{A}^{\Gamma _{0}}}(\xi _{h^{\ast }}^{Q})=\pi _{B^{\Gamma
}}(\xi _{h^{\ast }}^{Q})\overset{(\ref{second1})}{<}\pi _{B^{\Gamma }}(\xi
_{h^{\ast }}^{P})+\varepsilon +\frac{d}{2}\leq \pi _{\widehat{A}^{\Gamma
_{0}}}(\xi _{h^{\ast }}^{P})+\varepsilon +\frac{d}{2}\overset{(\ref{first})}{%
<}\pi _{\widehat{A}^{\Gamma _{0}}}(\xi _{h^{\ast }}^{Q}).
\end{equation*}%
which is a contradiction.
\end{proof}

\subsection{On quasiconvex real valued maps\label{sec52}}

\begin{proof}[Proof of Theorem \protect\ref{Volle1}]
By definition, for any $X^{\prime }\in L^{\prime },$ $R(X^{\prime
}(X),X^{\prime })\leq f(X)$ and therefore 
\begin{equation*}
\sup_{X^{\prime }\in L^{\prime }}R(X^{\prime }(X),X^{\prime })\leq f(X),%
\text{ }X\in L.
\end{equation*}%
Fix any $X\in L$ and take $\varepsilon \in \mathbb{R}$ such that $%
\varepsilon >0$. Then $X$ does not belong to the closed convex set $\left\{
\xi \in L:f(\xi )\leq f(X)-\varepsilon \right\} :=\mathcal{C}_{\varepsilon }$
(if $f(X)=+\infty $, replace the set $\mathcal{C}_{\varepsilon }$ with $%
\left\{ \xi \in L:f(\xi )\leq M\right\} ,$ for any $M$). By the Hahn Banach
theorem there exists a continuous linear functional that strongly separates $%
X$ and $\mathcal{C}_{\varepsilon }$, i.e. there exists $\alpha \in \mathbb{R}
$ and $X^{\prime }\in L^{\prime }$ such that 
\begin{equation}
X_{\varepsilon }^{\prime }(X)>\alpha >X_{\varepsilon }^{\prime }(\xi )\text{
for all }\xi \in \mathcal{C}_{\varepsilon }\text{.}  \label{100}
\end{equation}%
Hence:%
\begin{equation}
\left\{ \xi \in L:X_{\varepsilon }^{\prime }(\xi )\geq X_{\varepsilon
}^{\prime }(X)\right\} \subseteq (\mathcal{C}_{\varepsilon })^{C}=\left\{
\xi \in L:f(\xi )>f(X)-\varepsilon \right\}  \label{finitecase}
\end{equation}%
and%
\begin{eqnarray*}
f(X) &\geq &\sup_{X^{\prime }\in L^{\prime }}R(X^{\prime }(X),X^{\prime
})\geq R(X_{\varepsilon }^{\prime }(X),X_{\varepsilon }^{\prime }) \\
&=&\inf \left\{ f(\xi )\mid \xi \in L\text{ such that }X_{\varepsilon
}^{\prime }(\xi )\geq X_{\varepsilon }^{\prime }(X)\right\} \\
&\geq &\inf \left\{ f(\xi )\mid \xi \in L\text{ satisfying }f(\xi
)>f(X)-\varepsilon \right\} \geq f(X)-\varepsilon .
\end{eqnarray*}
\end{proof}

\begin{proposition}
\label{rapprR}Suppose $L$ is a lattice, $L^{\ast }=(L,\geq )^{\ast }$ is the
order continuous dual space satisfying $L^{\ast }\hookrightarrow L^{1}$ and $%
(L,\sigma (L,L^{\ast }))$ is a locally convex TVS. If $f:L\rightarrow 
\overline{\mathbb{R}}$ is quasiconvex, $\sigma (L,L^{\ast })$-lsc and
monotone increasing then%
\begin{equation*}
f(X)=\sup_{Q\in L_{+}^{\ast }\mid Q(\mathbf{1})=1}R(Q(X),Q).
\end{equation*}
\end{proposition}

\begin{proof}
We apply Theorem \ref{Volle1} to the locally convex TVS $(L,\sigma
(L,L^{\ast }))$ and deduce: 
\begin{equation*}
f(X)=\sup_{Z\in L^{\ast }\subseteq L^{1}}R(Z(X),Z).
\end{equation*}%
We now adopt the same notations of the proof of Theorem \ref{Volle1} and let 
$Z\in L$, $Z\geq 0$. Obviously if $\xi \in \mathcal{C}_{\varepsilon }$ then $%
\xi -nZ\in \mathcal{C}_{\varepsilon }$ for every $n\in \mathbb{N}$ and from (%
\ref{100}) we deduce: 
\begin{equation*}
X_{\varepsilon }^{\prime }(\xi -nZ)<\alpha <X_{\varepsilon }^{\prime
}(X)\quad \Rightarrow \quad X_{\varepsilon }^{\prime }(Z)>\frac{%
X_{\varepsilon }^{\prime }(\xi -X)}{n}\quad ,\quad \forall \,n\in \mathbb{N}
\end{equation*}%
i.e. $X_{\varepsilon }^{\prime }\in L_{+}^{\ast }\subseteq L^{1}$ and $%
X_{\varepsilon }^{\prime }\neq 0.$ Hence $X_{\varepsilon }^{\prime }(\mathbf{%
1})=E_{\mathbb{P}}[X_{\varepsilon }^{\prime }]>0$ and we may normalize $%
X^{\prime }$ to $X^{\prime }/X_{\varepsilon }^{\prime }(\mathbf{1})$.
\end{proof}

\end{document}